\documentclass[prb,eqsecnum,superscriptaddress,twocolumn]{revtex4}
\usepackage[dvips]{graphicx}
\usepackage{latexsym}
\usepackage{amsmath}
\preprint{cond-mat/0409470}
\begin{document}
\title{Competing orders II: the doped quantum dimer model}

\author{Leon Balents}
\affiliation{Department of Physics, University of California,
Santa Barbara, CA 93106-4030}

\author{Lorenz Bartosch}
\affiliation{Department of Physics, Yale University, P.O. Box
208120, New Haven, CT 06520-8120}

\affiliation{Institut f\"ur Theoretische Physik, Universit\"at
Frankfurt, Postfach 111932, 60054 Frankfurt, Germany}

\author{Anton Burkov}
\affiliation{Department of Physics, University of California,
Santa Barbara, CA 93106-4030}

\author{Subir Sachdev}
\affiliation{Department of Physics, Yale University, P.O. Box
208120, New Haven, CT 06520-8120}

\author{Krishnendu Sengupta}
\affiliation{Department of Physics, Yale University, P.O. Box
208120, New Haven, CT 06520-8120}

\date{September 19, 2004}

\begin{abstract}
We study the phases of doped spin $S=1/2$ quantum antiferromagnets
on the square lattice, as they evolve from paramagnetic Mott
insulators with valence bond solid (VBS) order at doping
$\delta=0$, to superconductors at moderate $\delta$. The interplay
between density wave/VBS order and superconductivity is
efficiently described by the quantum dimer model, which acts as an
effective theory for the total spin $S=0$ sector. We extend the
dimer model to include fermionic $S=1/2$ excitations, and show
that its mean-field, static gauge field saddle points have
projective symmetries (PSGs) similar to those of `slave' particle
U(1) and SU(2) gauge theories. We account for the non-perturbative
effects of gauge fluctuations by a duality mapping of the $S=0$
dimer model. The dual theory of vortices has a PSG identical to
that found in a previous paper (L. Balents {\em et al.},
cond-mat/0408329) by a duality analysis of bosons on the square
lattice. The previous theory therefore also describes fluctuations
across superconducting, supersolid and Mott insulating phases of
the present electronic model. Finally, with the aim of describing
neutron scattering experiments, we present a phenomenological
model for collective $S=1$ excitations and their coupling to
superflow and density wave fluctuations.
\end{abstract}

\maketitle

\section{Introduction}
\label{sec:intro}

In a previous paper\cite{psgbosons} (hereafter referred to as I),
we described the physics of two-dimensional superfluids in the
vicinity of a localization transition into a commensurate Mott
insulator. Our primary focus was on bosons on the square lattice,
and we demonstrated the role played by the projective
transformations (the PSG) of the vortices in the superfluid under
operations of the square lattice space group. The present paper
will extend our analysis to paired electron models on the square
lattice, with the aim of justifying applicability to the cuprate
superconductors.  A simple physical argument based on universality
was given in I clearly validating that analysis when the
superconducting state neighboring the Mott insulator is an
$s$-wave state with a large quasiparticle gap, and it was claimed
that the same description also holds for a similarly gapped
``strong pairing'' $d$-wave state.  In this paper, we back up this
claim with detailed microscopic analysis of a set of models that
incorporate local singlet formation. Specifically, we find that
most of the results of I can be applied essentially unchanged,
with the PSG determined by the particle density of Cooper pairs.
We will also discuss the extent to which our results can be
applied to ``weak pairing'' $d$-wave superconductors with gapless
nodal quasiparticles.

Most of our analysis here will be carried out in the context of
the quantum dimer model. This model was proposed by Rokhsar,
Kivelson, and Fradkin \cite{rk,fradkiv} as an effective theory of
the $S=0$ excitations of the insulating, paramagnetic phases of
square lattice antiferromagnets. For generic parameters, the
insulating dimer model has a ground state which breaks the
symmetry of the square lattice space group with the development of
valence bond solid (VBS) order \cite{rs,ssdimer,leung,ms}. The
dimer model was also extended \cite{rk,fradkiv} to allow for $S=0$
charged excitations: the Hilbert space was expanded to include
bosonic holes with density $\delta$, and it was argued that the
ground state is a superfluid at finite $\delta$ (the pairing
symmetry of this superconducting state is somewhat unclear in
these works, and will be clarified below). The doped quantum dimer
model is therefore ideally suited to our primary purpose here of
describing the evolution from the VBS Mott insulator to the
superfluid as a function of increasing $\delta$.

We will begin in Section~\ref{sec:dimer} by reviewing the
formulation \cite{fradkiv,curreac} of the quantum dimer model as a
compact U(1) gauge theory (with gauge field $\hat{\mathcal{A}}$)
in the presence of a background static matter field of charges
$\pm 1$ on the two sublattices of the square lattice. Upon doping,
the holes appear as two $S=0$ dynamic bose fields $b_{i \sigma}$
($i$ is a square lattice site index and the index $\sigma = \pm 1$
labels the sublattice upon which the boson primarily resides).
These bosons have charges $\sigma$ under the compact U(1) gauge
field $\hat{\mathcal{A}}$.

In Section~\ref{sec:spinful} we show how the above theory of the
spin $S=0$ sector can be extended to include $S=1/2$ fermionic
excitations $f_{is}$ ($s=\uparrow,\downarrow$ is a spin label).
These fermionic `spinons' also carry unit staggered charges under
the compact U(1) gauge field $\hat{\mathcal{A}}$. While these
fermionic excitations are gapped and confined in the undoped
insulator, the structure of our theory allows these fermions to
become gapless quasiparticles at nodal points in a possible
$d$-wave superconducting ground state at finite $\delta$: they
then become the conventional $S=1/2$ Bogoliubov quasiparticles of
this superconductor.  With this construction, we can directly
discuss the pairing symmetry of the superfluid states of the doped
dimer model. We will also consider collective $S=1$ `triplon'
excitations in Section~\ref{sec:triplon}.

So far we have mainly defined the degrees of freedom of our theory
on the lattice scale, but this generally does not specify its
ultimate ground state or the quantum numbers of its
quasiparticles. In particular, we are considering a strongly
interacting 2+1 dimensional compact U(1) gauge theory, and the
non-perturbative fluctuations of $\hat{\mathcal{A}}$ must be
treated with great care.  Generally, for the dimer models we
consider here, these fluctuations drive confinement at long
scales.  Nevertheless, it is useful to initially neglect gauge
fluctuations and to consider mean-field saddle points with a fixed
background $\hat{\mathcal{A}}$. Such saddle points describe
(unstable) `spin liquid' states, which may be considered the
progenitors of the superconducting and/or charge ordered Mott
phases which appear as ground states.  A subsequent treatment of
gauge fluctuations describes this evolutionary descent via
confinement or the Higgs mechanism. Remarkably, we will find in
Section~\ref{sec:conn-slave-part} that, at least at the level of
PSGs, the spin liquid saddle points of the dimer model are related
to particular (generalized) saddle points of earlier
slave-particle U(1) (Refs.~\onlinecite{ba,sr}) or SU(2)
(Ref.~\onlinecite{su2}) gauge theories.  The particular saddle
points appropriate to dimer models, however, are distinct from
those advocated as candidate spin liquid states in the slave
particle work.  They have appeared -- without much attention -- in
Wen's PSG classification of symmetric spin liquids \cite{wenpsg}
(the specific identifications of these spin liquids appear at the
end of Section~\ref{sec:conn-slave-part}). Our work thus resolves
one question posed in Ref.~\onlinecite{wenpsg}, as to how the
(putative) spin liquid states of the quantum dimer model fit into
this classification scheme. Beyond classification, the physical
interpretation we shall give to our dimer model saddle points, and
our subsequent non-perturbative treatment of gauge fluctuations,
lead us eventually to conclusions on the physical properties of
our theory which differ from those reached in
Refs.~\onlinecite{su2,su2long}.

There is one situation in which it is clear that the
$\hat{\mathcal{A}}$ gauge fluctuations are relatively innocuous.
This is the moderate $\delta$ state in which the $b_{i \sigma}$
are strongly condensed and the square lattice symmetry is fully
preserved: the condensate acts like a Higgs field which quenches
$\hat{\mathcal{A}}$ fluctuations. Such a state is a
superconductor, and we will show in Section~\ref{sec:superc-state}
that microscopic energetic considerations and the PSG constrain
the pairing symmetry of the superconductor; $d$- and $s$- wave
(and other) pairings emerge under suitable conditions.

So what is the fate of this putative $d$-wave superconductor as
$\delta$ is reduced? As noted above, we know from
Refs.~\onlinecite{rs,curreac} that strong $\hat{\mathcal{A}}$
fluctuations produce a VBS insulator at $\delta=0$, and so a
careful treatment of $\hat{\mathcal{A}}$ fluctuations must surely
become increasingly important as $\delta$ is reduced. In previous
work\cite{sr,vs,ps,mv} on U(1) slave particle and related
theories, the evolution between a finite $\delta$ $d$-wave
superconductor and a VBS insulator at $\delta=0$ was addressed in
a mean-field theory (similar to that just discussed above) but
which also allowed for breaking of the square lattice symmetry by
the development of density wave/VBS order coexisting with
superfluidity. This had the advantage of explicitly displaying the
evolution of the gapless fermionic nodal quasiparticle excitations
with decreasing $\delta$: the nodes present at larger $\delta$
disappeared (due to collisions in pairs at the (reduced) Brillouin
zone boundaries) at a critical doping $\delta=\delta_f$, larger
than the doping $\delta_c$ at which the Mott transition occurs.
For small enough $\delta<\delta_c$, a fully-gapped insulating VBS
state was obtained. However, this work treated gauge or VBS
fluctuations in a cavalier manner, and the VBS order was static
and likely overestimated in magnitude.

In the application to the cuprates, we have in mind that these
insulating or supersolid states obtained in our theory at finite
doping are models of the situation near doping $\delta=1/8$. There
is much experimental evidence ({\em e.g.\/} in
La$_{1.6-x}$Nd$_{0.4}$Sr$_x$CuO$_4$ in Ref.~\onlinecite{jtran})
that exactly at $\delta=1/8$ the superconductivity is strongly
suppressed, and perfect long-range order with a period of 4
lattice spacings develops (as is also the case in
Fig~\ref{fig:H=0.0625}).

In Section~\ref{sec:dimerdual}, we will address the $\delta$
driven evolution from a VBS insulator to a superconductor in an
approach which fully accounts for the non-perturbative effects of
the fluctuations of $\hat{\mathcal{A}}$. No previous analysis
\cite{sr,su2,vs,ps,mv,su2long,wenpsg} has accounted for strong
gauge fluctuations in this regime. However, we will only be able
to do this in the context of the $S=0$ model of
Section~\ref{sec:dimer}: non-zero $S$ excitations will be
neglected in Section~\ref{sec:dimerdual}. While this is probably
safe within the phases obtained therein, neglect of gapless nodal
$S=1/2$ fermionic excitations is likely invalid at the quantum
critical points. A full treatment which synthesizes the results of
Sections~\ref{sec:spinful} and~\ref{sec:dimerdual}, and thus
accounts for {\sl both} strong $\hat{\mathcal{A}}$ fluctuations
and gapless $S=1/2$ fermions at quantum critical points is not
available, and remains a challenging and important open problem.

The analysis of the $S=0$ sector in Section~\ref{sec:dimerdual}
is carried out by a duality mapping into a theory of vortices.
The resulting effective theory for the vortices has a structure
similar to the models considered in I, and can therefore be
similarly classified by an analysis of the action of the PSG on
the vortices. With the knowledge of the PSG, many of the results
of I can be directly applied here. We will also present in
Section~\ref{sec:dimermft} a mean-field analysis of the dual
vortex theory which results in a phase diagram of the dimer model
as a function of $\delta$. It is important to note that this dual
mean-field theory accounts for non-perturbative confining effects
of the $\hat{\mathcal{A}}$ fluctuations: these are built into the
duality mapping.

\section{Quantum dimer model}
\label{sec:dimer}

This section describes the doped quantum dimer model \cite{rk},
viewed as an effective theory of the spin $S=0$ sector of doped
quantum antiferromagnets. The resulting theory will be a compact
U(1) gauge theory \cite{fradkiv,curreac} coupled to two classes of
matter fields: ({\em i\/}) {\sl static} charges of $\pm 1$ on the
two sublattices, and ({\em ii\/}) {\sl dynamic} bosons of charges
$\pm 1$ with a total physical density $\delta$ (note: bosons of
both gauge charges contribute a positive number to the physical
density).

\begin{figure}
\centering
\includegraphics[width=3in]{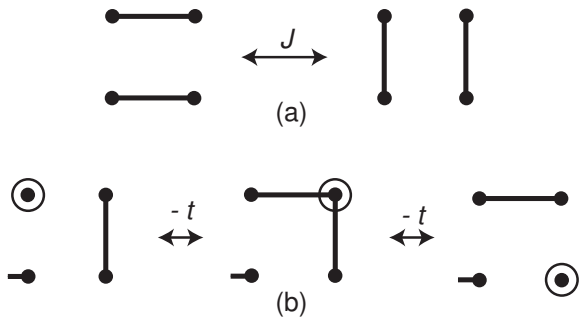}
\caption{Elementary moves of the quantum dimer model. ({\em a\/})
The resonance of two dimers around a plaquette, which occurs with
amplitude $J$ in Eq.~(\protect\ref{barhd}). ({\em b\/}) Hopping of
a hole with the amplitude $t$ term in Eq.~(\ref{barht}). The hole
(represented by the open circle) prefers to reside on one
sublattice. This hole is on the ``wrong'' sublattice (with high
energy cost) in the intermediate state above, which only acts as a
technical tool for effectively generating the same sublattice
hopping in the low energy limit. The constraint in
Eq.~(\protect\ref{constraint}) is obeyed on all sites in all the
states shown.} \label{dimer}
\end{figure}

The Hamiltonian of the quantum dimer model can be written as
\begin{equation}
\mathcal{H}_{qd} = \mathcal{H}_d + \mathcal{H}_h + \mathcal{H}_t
\label{hqd}
\end{equation}
The first term, $\mathcal{H}_d$ involves a diagonal energy and a
resonance of dimers on a plaquette, as illustrated in
Fig~\ref{dimer}a.
Following Ref.~\onlinecite{fradkiv}, and the notation of
Ref.~\onlinecite{curreac}, the dimers are represented by the
conjugate phase $\hat{\mathcal{A}}_{i\alpha}$ and integer number
$\hat{E}_{i\alpha}$ variables which reside on the links of the
lattice:
\begin{equation}
[\hat{\mathcal{A}}_{i\alpha},\hat{E}_{j\beta}] = i \delta_{ij}
\delta_{\alpha\beta}\;.
\end{equation}
The dimer number on a bond is $\eta_i \hat{E}_{i\alpha}$ oriented
to the right or the top ($\eta_i = +1$ or -1 on the two
sublattices). The Hamiltonian for dimers on a plaquette is
\begin{equation}
\mathcal{H}_d = \frac{V}{2} \sum_{i\alpha} \hat{E}_{i\alpha}^2 + J
\sum_i \cos \left( \epsilon_{\alpha\beta} \Delta_{\alpha}
\hat{\mathcal{A}}_{i \beta} \right) \label{barhd}
\end{equation}
Note that the co-efficient of the cosine has the ``wrong'' sign.  This
is a consequence of the microscopics of the electrons in the $t$-$J$
model, if one adopts the natural sign convention that a singlet dimer is
created by an electron pair-field operator (see
Eq.~(\ref{eq:pairfield})), and a given dimer covering corresponds to a
product of such operators acting on the vacuum.

The next two terms, $\mathcal{H}_h$ and $\mathcal{H}_t$, describe
the hopping of holes in the doped dimer model, as illustrated in
Fig~\ref{dimer}b. To maintain the description in terms of nearest
neighbor dimers, the holes must hop between sites on the same
sublattice. Consequently, the model has two species of holes, one
for each sublattice. This same-sublattice hole hopping model is
somewhat inconvenient to work with under the duality mapping, and
so we expand the Hilbert space of the model by additional high
energy states to allow us to express all hole motion in terms of
nearest neighbor hopping: we allow each hole species to
temporarily hop onto the ``wrong'' sublattice in a state which has
a large energy. This hole will quickly hop back onto the ``right''
sublattice, and the two hop process will have achieved the same
sublattice hopping we are seeking to model. This process is
illustrated in Fig~\ref{dimer}b; the same strategy was used by
Moessner {\em et al.} \cite{moessner} in their study of the doped
dimer model. We represent the two species of holes by the rotor
phases $\hat{\phi}_{i+}$, $\hat{\phi}_{i-}$ and the conjugate
number variables $\hat{n}_{i+}$, $\hat{n}_{i-}$. The commutation
relations are
\begin{equation}
[ \hat{\phi}_{i\sigma}, \hat{n}_{j\sigma'} ] = i \delta_{ij}
\delta_{\sigma\sigma'}
\end{equation}
where $\sigma = +,-$. These holes reside primarily on the $+$ or
$-$ sublattices and will have opposite gauge charges. The
potential energy of the holes is described by $\mathcal{H}_h$:
\begin{equation}
\mathcal{H}_h = \frac{U}{2} \sum_{i\sigma} \left(
\hat{n}_{i\sigma} - H - \sigma \eta_i W \right)^2 + \sum_{i \neq
j, \sigma} \Lambda_{ij} \hat{n}_{i \sigma} \hat{n}_{j \sigma} \;.
\label{eq:Hh}
\end{equation}
Here $U>0$ is the hole self repulsion, $2 H$ is the average hole
density preferred by the chemical potential (we are working at
fixed chemical potential, and the actual density, $\delta$, in the
ground state of the Hamiltonian may be different), $\eta_i = \pm
1$ on the two sublattices, and $2UW$ is the energy penalty for a
hole to be on the ``wrong'' sublattice. For $W
> 0$, the $+$ holes are preferred on the $\eta_i = 1$
sublattice, and the $-$ holes on the other. The off-site
interaction energy $\Lambda_{ij}$ is assumed to be repulsive, and
is required to stabilize insulating states away from zero doping.
We will discuss the case of general $\Lambda_{ij}$, although our
numerical results later will be restricted to short-range
$\Lambda_{ij}$; long-range Coulomb interactions lead only to minor
modifications. The total density of holes on site $i$ is given by
\begin{equation}
\delta_i = \sum_{\sigma} \langle \hat{n}_{i \sigma} \rangle \;.
\end{equation}
The nearest neighbor hopping Hamiltonian for the two hole species
is
\begin{equation}
\mathcal{H}_t = -t \sum_{i\alpha\sigma} \cos\left( \Delta_\alpha
\hat{\phi}_{i \sigma} + \sigma \hat{\mathcal{A}}_{i\alpha} \right)
\;. \label{barht}
\end{equation}

Finally, there is a local constraint, which commutes with all the
terms in the Hamiltonian,
\begin{equation}
\Delta_{\alpha} \hat{E}_{i\alpha} + \sum_{\sigma} \sigma
\hat{n}_{i\sigma} = \eta_i\;. \label{constraint}
\end{equation}
This constraint ensures that there is either a hole on each
lattice site or a dimer emerging from it. It also allows for a
configuration with two dimers emerging from that site if the extra
gauge charge is compensated by an additional hole from the
``wrong'' sublattice. This is the intermediate state in
Fig~\ref{dimer}b which compared with the other two states in that
figure requires the excitation energy $V/2 + U W$. We see that for
sufficiently large $V$ we can safely set $W$ equal to zero and
still pay a price for having a hole on the ``wrong'' sublattice.
Choosing both $V$ and $U$ sufficiently large also excludes
unphysical states with even larger excitation energies. Note that
all the states in Fig~\ref{dimer} obey Eq.~(\ref{constraint}).

We have already noted that the dimer resonance term in Eq.~(\ref{barhd})
has the opposite sign from that usually found for the Maxwell term in
lattice gauge theories.  We will see below that this can have
significant consequences for the physics, and hence the structure of the
dual theory.  Physically, the kinetic term in Eq.~(\ref{barhd}) favors
an average $\pi$ flux per plaquette in the gauge field
$\hat{\mathcal{A}}_{i\alpha}$.  The hole propagation term,
Eq.~(\ref{barht}), however, clearly allows holes to achieve lower
kinetic energy if this flux is expelled.  Since the gauge field is
dynamical, the system will choose for itself what flux penetrates each
plaquette on average, based on a balance of these two competing
energies, and as pointed out by Kivelson and Fradkin\cite{fradkiv}, the
flux will penetrate at $\delta=0$ and be expelled for large enough
$\delta$.  In a strongly confining phase, as occurs for $\delta=0$,
however, this flux has no physical significance, since the gauge field
is strongly fluctuating.

Mathematically, one may choose to view either the zero or $\pi$ flux
states as vacua, provided fluctuations of this flux are included.  In
particular, one may attempt to fix the sign ``problem'' ($J>0$) by shifting
the gauge field $\hat{\mathcal{A}}_{i\alpha}$ on certain links by $\pi$.
However, this has the effect of inducing a $\pi$ flux per plaquette in
the motion of the $\phi_{i \sigma}$ holes in Eq.~(\ref{barht}). If we
were to perform such a transformation, $\mathcal{H}_d$ and
$\mathcal{H}_t$ would be replaced by
\begin{equation}
\mathcal{H}_d^{\prime} = \frac{V}{2} \sum_i \hat{E}_{i\alpha}^2 -
J \sum_i \cos \left( \epsilon_{\alpha\beta} \Delta_{\alpha}
\hat{\mathcal{A}}_{i \beta} \right), \label{hd1}
\end{equation}
\begin{equation}
\mathcal{H}_t^{\prime} = -t \sum_{i\alpha\sigma} \cos\left(
\Delta_\alpha \hat{\phi}_{i \sigma} + \sigma
\hat{\mathcal{A}}_{i\alpha} + 2\pi g_{i \alpha} \right),
\label{ht1}
\end{equation}
and the holes now move in a magnetic field $g_{i \mu}$ which obeys
\begin{equation}
\epsilon_{\mu\nu\lambda} \Delta_{\nu} g_{i\lambda} = h \delta_{\mu
\tau} \label{curlg}
\end{equation}
with $h=1/2$.

For (most of) our purposes, it will generally not be advantageous to
make such a shift.  This is physically because our description of the
Mott transition approaches the critical point from the superconducting
side, in terms of the condensation of vortices of the superconducting
state.  As discussed above, in the large $\delta$ region of the dimer
model where superconductivity occurs, zero flux is favored.  Hence this
is a more natural point from which to embark upon our (dual)
investigation.

In the following section, however, in which we discuss quasiparticle
physics and pairing symmetry, we will consider the physical content of
this $\pi$-flux background.  Our intuition to expand around the zero
flux configuration is there confirmed, as we show that the $\pi$-flux
background is inconsistent with $d$-wave (or $s$-wave) pairing in the
superconductor, and applies instead to a yet more unconventional state.

\section{Adding fermions and spin to the dimer model}
\label{sec:spinful}

In the context of the cuprates, a shortcoming of the present
approach -- intrinsic to the dimer model -- is the neglect of
fermionic spin-carrying quasiparticles.  To understand the nature
of this difficulty more deeply, we discuss the results of a
natural extension of the dimer approach to include these $S=1/2$
fermionic excitations. The dimer model can also be extended to
include collective $S=1$ `triplon' excitations, but we defer
discussion of these to Section~\ref{sec:triplon}, until after we
have performed a duality analysis of the $S=0$ dimer model in
Section~\ref{sec:dimerdual}.

Readers mainly interested in the duality analysis of the $S=0$
dimer model at finite doping may skip ahead to
Section~\ref{sec:dimerdual}, and return to the present section
upon a second reading.

The strategy will be to first understand in more detail the nature
of the insulating states of this model when {\sl undoped}. Because
in this model at half-filling all the ``matter'' (holes and
fermions) is gapped, the Polyakov argument implies that the U(1)
gauge theory must be in a confining phase. Nevertheless, it is
instructive to consider formally the limit of weak gauge
fluctuations, which in practice applies for large $J$ -- of either
sign.  Then one may view the insulator as a gapped U(1) state plus
gauge fluctuations, the latter -- more specifically monopole
proliferation -- ultimately leading to confinement on long length
scales, and with it, density wave order.  This view has the
advantage that the superconducting state can be seen as a
``Higgs'' phase obtained from the U(1) state. Further, the
insulator can be characterized by a PSG, defining the manner in
which the physical symmetries are realized in the gauge theory.
Here we will consider two such PSGs, which correspond naturally to
the phases obtained for large positive and negative $J$.

We can better understand in this manner the nature of the
superconducting and insulating states in this model.  We will see
that the former should be understood as a ``molecular'' or
``strong pairing'' d-wave phase.  The latter can be seen to arise
from either of two natural underlying U(1) spin liquid states.
These two U(1) states are distinct, and characterized by different
PSGs.  They are, however, both unstable at low energies to the
various charge ordered states discussed herein.

We proceed with a multi-pronged attack.  First, we describe the
symmetry of the bosonic dimer+hole Hamiltonian of
Sec.~\ref{sec:dimer}.  We consider the two distinct cases in which
the ground state has an average zero or $\pi$ gauge flux per
plaquette. These two cases give rise to distinct PSGs which
describe different putative insulating states.  With this in hand,
we discuss the nature of the pairing symmetry of superconducting
states obtained from this model.  Next, we will augment the
dimer+hole Hilbert space by allowing each site to contain an
unpaired electron (or spin), in addition to the end of a dimer or
the hole allowed up to now. Physical arguments give the general
form of the Hamiltonian with such unpaired electrons included.
This form is further restricted by combining the PSGs of the
dimer+hole model and the requirement that physical electrons
transform properly.  We discuss how this analysis is related to
slave particle formulations of Mott insulating states and Wen's
analysis \cite{wenpsg} of their PSGs. With the full form of the
dimer+hole+electron Hamiltonian determined, we confirm the
symmetry analysis of the superconducting states by a direct
investigation of the effective Bogoliubov-deGennes quasiparticle
Hamiltonian.

\subsection{Projectively realized symmetries of the doped dimer model}
\label{sec:proj-real-symm}

It is clear that, even without including fermionic excitations,
the dimer model in Sec.~\ref{sec:dimer} already requires a
non-trivial projective realization of certain symmetry operations
of the space group.  This is so because the staggered static gauge
charges (and the staggered potential term $W$ in
Eq.~(\ref{eq:Hh})) na\"ively break lattice symmetries. To overcome
this, unit translations and $\pi/2$ rotations about a dual lattice
site must be accompanied by the transformations $\hat{n}_+
\leftrightarrow \hat{n}_-$, $\hat\phi_+ \leftrightarrow
\hat\phi_-$, $\hat{\cal A} \rightarrow -\hat{\cal A}$,
$\hat{E}\rightarrow -\hat{E}$ to leave the dimer model invariant.
Actually, it will prove useful to discuss projective symmetries
with respect to a fixed average background U(1) flux. This will
make contact with Wen's notion of projective symmetry analysis of
slave particle saddle points.  We will consider the two cases of
zero and $\pi$ average flux per plaquette.

For compactness, we will in this section denote the boson
operators by
\begin{equation}
  \label{eq:bosedef}
  b_{i\sigma} = e^{-i\hat\phi_{i\sigma}}.
\end{equation}
Further, it is sometimes helpful to take the limit
$W\rightarrow\infty$, corresponding to a ``microscopic''
dimer+hole model in which two holes of course cannot occupy the
same site.  In this case the type $1$ holes live only on the $A$
sublattice, and conversely the type $2$ holes live only on the
opposite $B$ sublattice.  One then defines a single boson field
via
\begin{equation}
  \label{eq:bslave}
  b_i = \left\{
    \begin{array}{cc} e^{-i\hat\phi_{i+}} & \mbox{for $i \in$ A sublattice} \\
      e^{-i\hat\phi_{i-}} & \mbox{for $i \in$ B sublattice}
  \end{array} \right. ,
\end{equation}
i.e. $b_i=b_{i\sigma}$ if $i$ is on the $\sigma$ sublattice.  For
$W\rightarrow\infty$, all finite energy states can be described in
terms of the $b_i$ operator.  For $W<\infty$, this is not the
case, but $b_i$ defined as above is still useful and represents
properly the low energy hole degrees of freedom.

\subsubsection{Zero flux background}
\label{sec:zero-flux-background}

First, and simplest, is the case of zero flux, the na\"ive choice
for $J<0$ (note that which saddle point is chosen by the system is
however a dynamical question, which will in general depend upon
all the degrees of freedom, and may vary as other parameters in
the Hamiltonian are varied).  In this case, one can easily see
that the only non-trivial {\sl required} aspect of transformations
of the bosons is the interchange of the two boson flavors -- and
associated change of the sign of gauge fields -- for space group
operations that interchange the two sublattices.  Specifically,
these are unit translations (or generally translations by an odd
number of total lattice units), $\pi/2$ rotations about a dual
lattice site, and inversions through a row/column of the dual
lattice.  However, there is a mathematical freedom to compose each
space group operation with a global U(1) phase rotation.  That is,
for an operation which preserves the two sublattices, one may take
a priori
\begin{equation}
  \label{eq:glob}
  b_{i\sigma} \rightarrow b_{G(i)\sigma}e^{i\chi(G)}.
\end{equation}
Similarly, for an operation that interchanges the two sublattices,
\begin{equation}
  \label{eq:glob2}
  b_{i\sigma} \rightarrow b_{G(i),-\sigma}e^{i\chi(G)}\;.
\end{equation}
Using the definition Eq.~(\ref{eq:bslave}), both these equations
are replaced simply by
\begin{equation}
  \label{eq:glob3}
  b_{i} \rightarrow b_{G(i)}e^{i\chi(G)}.
\end{equation}
Here $G(i)$ is the site $i$ is mapped to under the space group
operation $G$, and $\chi(G)$ is the phase, which we take as {\sl
the
  same} for both boson flavors, and can be chosen differently for each
$G$.  Note that this phase rotation is not a U(1) gauge rotation
(which would be opposite for the two boson flavors) unless
$e^{i\chi(G)}=\pm 1$.  Since the hole fields are not themselves
physical, it is not a priori obvious which value of $\chi(G)$ is
the {\sl correct} one.

An important constraint on $\chi(G)$ is that any sequence of space
group operations which compose to the trivial identity operation
in the original space group should compose to a pure gauge U(1)
transformation in the PSG.  The square of an inversion or the
fourth power of a $\pi/2$ rotation must be a pure U(1) gauge
transformation.  Hence the phase $e^{4i\chi(I)}= 1$ if $I$ is an
inversion, while $e^{8i\chi(R)}= 1$ if $R$ is a $\pi/2$ rotation.
Other similar constraints can be applied with more detailed
considerations.

For the doped dimer model, we can further refine this by using the
physical interpretation of the hole operators.  Consider the
action of the physical singlet pair field operator on a bond:
\begin{equation}
  \label{eq:pairfield}
  \Psi_{i\alpha} = c_{i,s}^\dagger \epsilon_{ss'}
  c_{i+\hat{e}_\alpha,s'}^\dagger,
\end{equation}
with $\alpha=x,y$. Clearly, this operator, when acting on a state
in which the bond in consideration is occupied by two holes,
annihilates the holes and creates a dimer.  Hence we require
\begin{equation}
  \label{eq:pfrequire}
  \Psi_{i\alpha} = \mu_\alpha b_i b_{i+\hat{e}_\alpha} e^{i\eta_i
    \hat{\cal A}_{i\alpha}} .
\end{equation}
Here we have introduced two unknown coefficients, $\mu_x,\mu_y$
(these are assumed independent of $i$, as can be argued is
generally true).  To determine them, we use the known
transformation properties of the pair field defined from
Eq.~(\ref{eq:pairfield}).  First, it is invariant under inversion
through the row/column of the dual lattice dividing the two sites
of the pair.  This implies $e^{2i\chi(I)}=1$ (actually this is
true for all inversions).  Then $e^{i\chi(I)}=\pm 1$, and for such
values the phase rotation in Eq.~(\ref{eq:glob2}) {\sl is} a gauge
U(1) rotation.  Thus we can always choose
\begin{equation}
  \label{eq:invreq}
  e^{i\chi(I)} = 1.
\end{equation}
Second, consider a $\pi/2$ rotation.  Under rotation around site
$i$, Eq.~(\ref{eq:pairfield}) implies $\Psi_{ix} \rightarrow
\Psi_{iy}$, $\Psi_{iy} \rightarrow \Psi_{i-\hat{e}_x,x}$, while
from Eq.~(\ref{eq:pfrequire}), one finds
\begin{eqnarray}
  \label{eq:roti}
  \Psi_{ix} & \rightarrow & e^{2i\chi(R)} \frac{\mu_x}{\mu_y}
  \Psi_{iy}, \\
  \Psi_{iy} & \rightarrow & e^{2i\chi(R)} \frac{\mu_y}{\mu_x}
  \Psi_{i-\hat{e}_x,x}.
\end{eqnarray}
Hence
\begin{equation}
  \label{eq:rotreq}
  e^{2i\chi(R)} = \frac{\mu_x}{\mu_y} =\pm 1.
\end{equation}
No further requirements on $\chi(R)$ and $\mu_\alpha$ follow from
symmetry alone.  The two choices in Eq.~(\ref{eq:rotreq}) simply
represent different possible PSGs.  Note that this choice is
directly related to the pairing symmetry of the superconducting
state.  If $\mu_x=\mu_y$, then a uniform condensate $\langle b_i
\rangle = {\rm constant}$ represent an $s$-wave paired state,
while for $\mu_x=-\mu_y$, the same condensate represents
$d_{x^2-y^2}$-wave pairing.   More generally, using
Eq.~(\ref{eq:pfrequire}), one can show
\begin{equation}
  \label{eq:psi4}
  \frac{\Psi_{iy} \Psi_{i+\hat{e}_x,y}}{\Psi_{ix} \Psi_{i+\hat{e}_y,x}}
  = \left(\frac{\mu_y}{\mu_x}\right)^2 e^{-i \eta_i
    \epsilon_{\alpha\beta} \Delta_\alpha \hat{\mathcal{A}}_{i\beta}},
\end{equation}
where the curl in the exponential on the right hand side is the
gauge flux through the plaquette with site $i$ at the lower-left
corner. Hence for the zero background flux states we consider
in this subsubsection, using Eq.~(\ref{eq:rotreq}), one sees that
\begin{equation}
  \label{eq:psi4a}
  \frac{\Psi_{iy} \Psi_{i+\hat{e}_x,y}}{\Psi_{ix} \Psi_{i+\hat{e}_y,x}}
  = 1.
\end{equation}
This allows for both $s$ and $d$-wave pairing.

\subsubsection{$\pi$ flux background}
\label{sec:pi-flux-background}

Let us now turn to the more complicated case in which half a
quantum of background gauge flux pierces each plaquette of the
direct lattice. We write
\begin{equation}
  \label{eq:bgflux}
  \hat{A}_{i\alpha} \rightarrow \overline{A}_{i\alpha} + \hat{A}_{i\alpha},
\end{equation}
where $\overline{A}_{i\alpha}$ is a classical c-number background
vector potential representing $\pi$ flux per plaquette.  An
extremely convenient and symmetrical choice for this section is
\begin{equation}
  \label{eq:bgvp}
  \overline{\cal A}_{ix} = -\overline{\cal A}_{iy} = \frac{\pi}{4}\eta_i.
\end{equation}
One can readily see that, with this choice, the Hamiltonian
remains invariant under the same translational operations as for
zero background flux, i.e. for a unit translation,
\begin{equation}
  \label{eq:Trans}
  \hat{T}_\alpha: \qquad b_{i,\sigma} \rightarrow
  b_{i+\hat{e}_\alpha,\overline{\sigma}},
\end{equation}
and $\hat{\cal A}_{i\beta} \rightarrow - \hat{\cal
  A}_{i+\hat{e}_\alpha,\beta}$ etc.  Rotations about dual lattice sites,
and inversions are also unchanged.  Rotations around direct
lattice sites are, however, non-trivial.  For a $\pi/2$ rotation
about a direct lattice site, one requires
\begin{eqnarray}
  \label{eq:Rotpi}
  b_{i1} & \rightarrow & e^{i\chi(R)}\zeta_i b_{R(i)1}, \\
  b_{i2} & \rightarrow & e^{i\chi(R)}\zeta^*_i b_{R(i)2},
\end{eqnarray}
where
\begin{equation}
  \label{eq:zetadef}
  \zeta_i = \left\{
    \begin{array}{cl}
      1 & \mbox{for $(i_x,i_y)=(0,0)\, ({\rm mod} 2)$}, \\
      -i & \mbox{for $(i_x,i_y)=(1,0)\, ({\rm mod} 2)$}, \\
      -1 & \mbox{for $(i_x,i_y)=(1,1)\, ({\rm mod} 2)$}, \\
      i & \mbox{for $(i_x,i_y)=(0,1)\, ({\rm mod} 2)$}.
    \end{array} \right.
\end{equation}
Note that, since $\zeta_i$ is real on the $A$ sublattice, one has
simply
\begin{equation}
  \label{eq:rotpib}
  b_i \rightarrow e^{i\chi(R)} \zeta_i^* b_{R(i)}.
\end{equation}

We can now repeat the symmetry analysis done above for zero
background flux.  As before, assuming the form of
Eq.~(\ref{eq:pfrequire}), since the transformation law for
inversions has not changed, one still obtains
Eq.~(\ref{eq:invreq}).  Under rotations, however, one obtains the
conditions
\begin{equation}
  \label{eq:fluxconds}
  e^{2i\chi(R)} = -i \frac{\mu_x}{\mu_y} =  \pm 1.
\end{equation}
Note the important factor of $i$ in Eq.~(\ref{eq:fluxconds})
compared to Eq.~(\ref{eq:rotreq}).  This implies very different
pairing states if the bosons condensed in the $\pi$ flux
background.  For instance, if the holons condense uniformly,
$\langle b_i \rangle = {\rm const.}$, then $\Psi_{iy} = \mp i
\Psi_{ix}={\rm const.}$.  This is incompatible with either $d$- or
$s$-wave pairing, and indeed with rotational invariance (in a
$d$-wave superconductor, rotational symmetry is unbroken for
gauge-invariant observables).

One way to see this is to consider the ratio in
Eq~(\ref{eq:psi4}), which becomes in this case
\begin{equation}
  \label{eq:psi4b}
  \frac{\Psi_{iy} \Psi_{i+\hat{e}_x,y}}{\Psi_{ix} \Psi_{i+\hat{e}_y,x}}
  = -1,
\end{equation}
irrespective of the spatial pattern of boson condensation.  In more
conventional terms, the superconducting states with holons condensed in
the $\pi$-flux background have a pair wavefunction of $(1\pm
i)s_{x^2+y^2} + (1\mp i)d_{x^2-y^2}$ form.  Again it is clear that such
a complex linear combination of extended $s$- and $d$-wave pairing
breaks fourfold rotational symmetry.

\subsection{Fermions}
\label{sec:fermions}

Having understood the possible PSGs of the dimer+boson model, we
are prepared to discuss fermionic excitations.  In principle,
there are many distinct ways in which the fermions might transform
consistent with the above PSGs.  There is, however, clearly a
physically preferred and simple choice.  We proceed by allowing a
site to be occupied not only by a hole and an end of a dimer, but
also by an unpaired electron.  This enlargement of the Hilbert
space is represented by a spin-$1/2$ fermion mode on each site,
described by canonical fermion creation/annihilation operators
$f^\dagger_{is}, f^{\vphantom\dagger}_{is}$ ($s$ is a spin-$1/2$
index). The original model is to be viewed as the fermion vacuum.
Clearly, the electron annihilation operator may act on a site with
such an unpaired electron to create a hole and remove the fermion.
Hence we require
\begin{equation}
  \label{eq:electrondef}
  c_{is} \sim b_i^\dagger f^{\vphantom\dagger}_{is}+ \cdots
\end{equation}
the ellipses are included to emphasize that of course
microscopically the electron annihilation operator can create
other states.  For instance, it may act upon a site which is part
of a dimer to annihilate the dimer, create a hole on the site
acted upon and create a fermion on the other site shared by the
dimer; such terms would be of the form
\begin{equation}
  \label{eq:electrondef2}
  c_{is} \sim \cdots b_i^\dagger e^{i \hat{\mathcal{A}}_{i \alpha}} \epsilon_{ss'}
  f^{\dagger}_{i+\hat{e}_\alpha,s'}+ \cdots\;.
\end{equation}
Such additional terms are of quantitative importance in analyzing
a microscopic dimer+hole+fermion model. In any case, all such
terms are such that the constraint in Eq.~(\ref{constraint}) is
now expanded to
\begin{equation}
\Delta_{\alpha} \hat{E}_{i\alpha} + \sum_\sigma \sigma
\hat{n}_{i\sigma} + \eta_i \sum_s f^{\dagger}_{is}
f^{\vphantom\dagger}_{is}= \eta_i. \label{constraint2}
\end{equation}
This constraint makes it clear that the $f_{is}$ fermions carry
staggered U(1) gauge charge.

The single term in Eq.~(\ref{eq:electrondef}), however, suffices
to determine the transformation properties of the $f_{is}$
operators, from the requirement that the physical electron
operator $c_{is}$ is a scalar. First, we see that the $f_{is}$
fermion carries a staggered gauge charge of opposite sign on the
two sublattices. Next, we note that with zero background flux,
$b_i$ is a scalar apart from the $\chi(G)$ factor discussed in the
previous section. Hence
\begin{equation}
  \label{eq:zerofluxft}
  f_{is} \rightarrow e^{i\chi(G)} f_{G(i)s}, \qquad \mbox{for zero
  background flux}.
\end{equation}
With the $\pi$ flux background, Eq.~(\ref{eq:zerofluxft}) holds
for transformations other than the $\pi/2$ rotation about a direct
lattice site.  For this, one has
\begin{equation}
  \label{eq:pifluxfrot}
  f_{is} \rightarrow e^{i\chi(R)} \zeta^*_i f_{R(i)s}, \qquad
  \mbox{for $\pi$ background flux}.
\end{equation}

At this point we note that the similarity between
Eq.~(\ref{eq:electrondef}) and the means of introducing slave
particle ``holon'' and ``spinon'' operators in the U(1) mean field
theory of the $t-J$ model is not coincidental.  Indeed, the
universal aspects of this discussion can all be recovered as
appropriate saddle points of that approach.  They can also be
recovered from the SU(2) mean field theory, the two approaches
being interchangeable in this instance, as will be returned to
below.

Having discussed the transformation properties of the fermions, we
proceed to construct their Hamiltonian -- assuming initially zero
background field.
\begin{figure}
\centering
\includegraphics[width=2.7in]{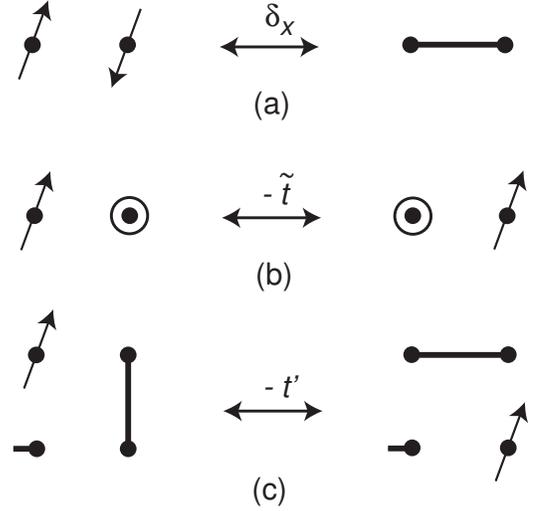}
\caption{Extension of Fig~\ref{dimer} to include $S=1/2$
excitations in the dimer model. ({\em a\/}) Two $S=1/2$ fermions
forming a singlet valence bond, representing Eq.~(\ref{eq:Hf}).
({\em b\/}) A holon and a $S=1/2$ fermion exchanging positions,
representing the first term in Eq.~(\ref{eq:Hfhop}). ({\em c\/})
Motion of a $S=1/2$ fermion involving rearrangement of valence
bonds, representing the second term in Eq.~(\ref{eq:Hfhop}).}
\label{fermion}
\end{figure}
Physically, the simplest allowed process is one
in which a dimer converts to a singlet fermion pair (see
Fig~\ref{fermion}):
\begin{eqnarray}
  \label{eq:Hf}
  {\cal H}_{f}^{(1)} & = & v\sum_{i} f_{is}^\dagger
  f_{is}^{\vphantom\dagger} \\ && +\sum_{i\alpha} \delta_\alpha
  e^{-i\eta_i\hat{\cal
      A}_{i\alpha}} f_{is}^\dagger \epsilon_{ss'}
  f_{i+{\hat{e}_\alpha},s'}^\dagger + {\rm H.c.}. \nonumber
\end{eqnarray}
Here $v$ represents the energy cost to introduce a single fermion,
and the sums over the repeated spin indices $s$, $s'$ are
implicit. Taking $v\rightarrow \infty$ recovers the original dimer
model. The coefficients $\delta_x,\delta_y$ appear as ``pair
fields'' for the $f_{is}$ fermions.  They are constrained by
symmetry to satisfy
\begin{equation}
  \label{eq:pazf}
  \delta_y = e^{-2i\chi(R)} \delta_x=\pm \delta_x, \qquad \mbox{for zero background
    flux}.
\end{equation}

We would also like to allow for hopping processes for the
fermions.  A fermion can hop between neighboring sites by
exchanging places with a hole, or a fermion can hop between second
neighbor sites if an intermediate neighboring site is connected to
the destination site by a dimer, the dimer reconnecting after the
hop to the origin site.  These processes are represented by
\begin{eqnarray}
  \label{eq:Hfhop}
&&   {\cal H}_{f}^{(2)}  =  - \sum_{i\alpha} \tilde{t}b_i^\dagger
  b_{i+\hat{e}_\alpha}^{\vphantom\dagger}
  f_{i+\hat{e}_\alpha,s}^{\dagger} f_{is}^{\vphantom\dagger} + {\rm H.c.}
  \\
  & & - \sum_i t'\left[ f_{is}^\dagger
  f_{i+w,s}^{\vphantom\dagger}\Gamma_{i,i+w}
 +f_{is}^\dagger
  f_{i+\overline{w},s}^{\vphantom\dagger}
  \Gamma_{i,i+\overline{w}}  + {\rm H.c.} \right]
  \nonumber
\end{eqnarray}
where
\begin{eqnarray}
  \label{eq:gammadefs}
  \Gamma_{i,i+w} & = & e^{-i\eta_i(\hat{\cal
      A}_{ix} + \hat{\cal A}_{i+\hat{e}_x,y})} + e^{-i\eta_i(\hat{\cal
      A}_{iy} + \hat{\cal A}_{i+\hat{e}_y,x})}, \\
  \Gamma_{i,i+\overline{w}} & = & e^{-i\eta_i(\hat{\cal
      A}_{ix} - \hat{\cal A}_{i+\overline{w},y})} + e^{-i\eta_i(\hat{\cal
      A}_{i-\hat{e}_y,x} - \hat{\cal A}_{i-\hat{e}_y,y})},\nonumber
\end{eqnarray}
and $w=\hat{e}_x+\hat{e}_y$, $\overline{w}=\hat{e}_x-\hat{e}_y$.
The complicated-looking $\Gamma$ factors are simply the uniform
superposition of the gauge connections (exponential of the
discrete line integral of the gauge field) taken along the two
shortest paths connecting diagonal sites.  The equal superposition
of both paths is required to preserve inversion symmetry through
the $(11)$ and $(1\overline{1})$ axes of the square lattice.  We
give these $t'$ terms primarily for completeness, as they will not
have an important role in much of what follows.

By examination of Eqs.~(\ref{eq:Hf}),~(\ref{eq:Hfhop}), one sees
that the $f_{is}$ fermions hop unassisted (i.e. without
accompanying holes) only between sites of the same sublattice, and
experience pair fields on links connecting sites of opposite
sublattices.  This is indeed required by the staggered gauge
charge of the $f_{is}$ variables.  It is thus sometimes convenient
to define new variables via a particle-hole transformation:
\begin{eqnarray}
  \label{eq:spinons}
  d_{is} & = & \left\{
    \begin{array}{rl}
      f_{is} & \mbox{for $\eta_i=+1$}, \\
      \epsilon^{\vphantom\dagger}_{ss'}f_{is'}^\dagger & \mbox{for
        $\eta_i=-1$}
    \end{array} \right.
\end{eqnarray}
These $d_{is}$ variables have a constant unit gauge charge on
every site and transform as global SU(2) spinors.  They should
thus be identified as ``spinons''.  In these variables, one finds
that the constraint in Eq.~(\ref{constraint2}) now takes the form
\begin{equation}
\Delta_{\alpha} \hat{E}_{i\alpha} +
b^{\dagger}_{i+}b^{\vphantom\dagger}_{i+}-b^{\dagger}_{i-}b^{\vphantom\dagger}_{i-}
+ \sum_s d^{\dagger}_{is} d^{\vphantom\dagger}_{is}= 1,
\label{constraint3}
\end{equation}
while the Hamiltonian ${\cal
  H}_f={\cal H}_{f}^{(1)}+{\cal H}_{f}^{(2)}$ has the form
\begin{eqnarray}
  \label{eq:Hfd}
  {\cal H}_f  & = &  v\sum_{i} \eta_i d_{is}^\dagger
  d_{is}^{\vphantom\dagger} +\sum_{i\alpha} t_{i\alpha} e^{-i\hat{\cal
      A}_{i\alpha}} d_{is}^\dagger
  d_{i+{\hat{e}_\alpha},s}^{\vphantom\dagger} + {\rm H.c.} \nonumber \\
  && + \tilde{t} \Big[\sum_{i\in A,\alpha} b_{i+{\hat{x}_\alpha}}^\dagger
  b_i^{\vphantom\dagger} d_{is}^\dagger \epsilon_{ss'}
  d_{i+{\hat{x}_\alpha},s'}^\dagger\nonumber \\ && + \sum_{i\in B,\alpha} b_i^\dagger
  b_{i+{\hat{x}_\alpha}}^{\vphantom\dagger} d_{is}^\dagger \epsilon_{ss'}
  d_{i+{\hat{x}_\alpha},s'}^\dagger+ {\rm H.c.}\Big],
\end{eqnarray}
and we have dropped the $t'$ terms for brevity.  Here the pairing
terms for the $f_{is}$ fermions have become hopping amplitudes:
\begin{equation}
  \label{eq:deltatot}
  t_{i\alpha} = \left\{
    \begin{array}{rl}
      \delta_{i\alpha} & \mbox{for $\eta_i=+1$}, \\
       -\delta_{i\alpha}^* & \mbox{for
        $\eta_i=-1$\;.}
    \end{array} \right.
\end{equation}

From Eq.~(\ref{eq:Hfd}) one can clearly see that, if no holes are
present (or if the $b_i$ excitations are gapped), then the
$\tilde{t}$ term can be neglected, and the spinons $d_{is}$ obey a
simple tight-binding model.  They indeed carry unit gauge charge,
since they are minimally coupled to $\hat{\mathcal{A}}_{i\alpha}$.
We expect that the confining effects of $\hat{\mathcal{A}}$
fluctuations will produce VBS order in the insulator, and also
bind the spinons into pairs which result in collective $S=0$ and
$S=1$ excitations. Alternatively, the spinons can also confine
with holons to create charged $S=1/2$ excitations with the same
quantum numbers as an electron.

In contrast, in the superconducting state (see
Section~\ref{sec:superc-state} below) the $b_{i}$ are condensed,
and then ${\cal H}_f$ has the structure of the Bogoliubov theory
of a superconductor, with the $d_{is}$ acting as the $S=1/2$
Bogoliubov quasiparticles; the $b_i$ condensate also acts like a
Higgs field which quenches the $\hat{\mathcal{A}}$ fluctuations.
Indeed, if the $b_i$ condensate is strong enough, there is nothing
in Eq.~(\ref{eq:Hfd}) which prevents the appearance of gapless,
nodal, Bogoliubov quasiparticles in a $d$-wave superconducting
state (see also Section~\ref{sec:superc-state}). In a
superconductor which preserves all lattice symmetries, the nodal
quasiparticles will appear above a critical doping
$\delta=\delta_f$ as a quartet near the center of the Brillouin
zone in a strong-pairing to weak-pairing transition \cite{demelo}.
On the other hand, the superconductor could also have broken
lattice symmetries ({\em i.e.\/} it is a supersolid) and then the
transition at $\delta=\delta_f$ is associated with the pairwise
collision of nodal points at the reduced Brillouin zone boundaries
\cite{vs,ps,granath}.

Finally, it is useful to rewrite the expression for the physical
electron field, Eq.~(\ref{eq:electrondef}), in the spinon
variables:
\begin{eqnarray}
  \label{eq:electron}
  c_{is} & \sim & \left\{
    \begin{array}{rl}
      b_i^\dagger d^{\vphantom\dagger}_{is} & \mbox{for $\eta_i=+1$}, \\
       -b_i^\dagger \epsilon_{ss'}d_{is'}^\dagger & \mbox{for
        $\eta_i=-1$}
    \end{array} \right.
\end{eqnarray}
At low energies, or in the $W\rightarrow \infty$ limit, this can
be rewritten as
\begin{equation}
  \label{eq:electron2}
  c_{is} \sim b_{i+}^\dagger d^{\vphantom\dagger}_{is} -
  b_{i-}^\dagger \epsilon_{ss'} d_{is'}^\dagger,
\end{equation}
since the first and second terms contribute predominantly on the
$A$ and $B$ sublattices, due to the preferred locations of the two
hole species.

How does this change for the $\pi$ flux background?  The needed
form is obtained simply from Eqs.~(\ref{eq:Hf})-(\ref{eq:Hfd}) by
shifting according to Eq.~(\ref{eq:bgflux}).  One notes that the
$d_{is}$ spinons, being minimally coupled to the gauge field,
indeed propagate in this case according to a $\pi$-flux hopping
model. Specifically, the shift implies
\begin{eqnarray}
  \label{eq:pishift}
  \delta_x & \rightarrow & \delta^{(\pi)}_x = \delta_x e^{-i\pi/4}, \\
  \delta_y & \rightarrow & \delta^{(\pi)}_y = \delta_y e^{+i\pi/4},
\end{eqnarray}
whence
\begin{eqnarray}
  \label{eq:pihop}
  t_{ix} & \rightarrow & t^{(\pi)}_{ix} = \eta_i \delta_x e^{-i\eta_i \pi/4},
  \\
  t_{iy} & \rightarrow & t^{(\pi)}_{iy} = \eta_i \delta_x
  e^{-2i\chi(R)} e^{+i\eta_i \pi/4},
\end{eqnarray}
for real $\delta_x$.  Further, it is interesting to note in this
case that the $t'$ terms in Eq.~(\ref{eq:Hfhop}) {\sl vanish} with
the replacement $\hat{A}_{i\alpha} = \overline{A}_{i\alpha}$ due
to destructive interference of the two hopping paths around the
plaquette containing the two diagonal sites in question.  Indeed,
one can show from symmetry arguments alone that a diagonal hopping
term is inconsistent with the $\pi$-flux PSG, using
Eq.~(\ref{eq:pifluxfrot}) for rotations and
Eq.~(\ref{eq:zerofluxft}) for inversions.

\subsection{Connection to slave particle theories}
\label{sec:conn-slave-part}

We comment briefly on the connection of the above discussion to
the U(1) and SU(2) slave particle approaches to the $t-J$ model.
The dimer model Hamiltonian above can be regarded as a theory of
fluctuations around particular mean-field states of these
approaches. This potentially allows one to connect to other
gauge-theoretic scenarios and phases which are not within the
purview of the dimer model.  First consider the U(1)
approach.\cite{ba,sr} One introduces microscopically holon and
spinon fields via
\begin{equation}
  \label{eq:shu1}
  c_{is} = b_i^\dagger f_{is},
\end{equation}
with the constraint $b_i^\dagger
b_i^{\vphantom\dagger}+f_{is}^\dagger
f_{is}^{\vphantom\dagger}=1$.  Eq.~(\ref{eq:shu1}) should be
compared with Eq.~(\ref{eq:electrondef}).  Note that here there is
only a single species of holon on each site, so we must compare
directly to the dimer model in the large $W$ limit.  In the path
integral Lagrangian, this constraint is implemented by a Lagrange
multiplier $A_{i0}$:
\begin{equation}
  \label{eq:llm}
  {\cal L}_0 = -\sum_i i A_{i0}(b_i^\dagger
  b_i^{\vphantom\dagger}+f_{is}^\dagger f_{is}^{\vphantom\dagger}-1).
\end{equation}
The electron hopping ($t$) and Heisenberg ($J$) terms of the $t-J$
model are decoupled with matrix fields $Q_{ij},\overline{Q}_{ij},
\Delta_{ij}$ leading to quadratic terms in the bosons and fermions
of the form
\begin{eqnarray}
  \label{eq:ldecu1}
  {\cal L}_1 & = & \sum_{ij} \Big(Q_{ij} b_i^\dagger b_j^{\vphantom\dagger}
  + \overline{Q}_{ij} f_{is}^\dagger f_{js}^{\vphantom\dagger} +
  \Delta_{ij} f_{is}^\dagger \epsilon_{ss'} f_{js'}^\dagger + {\rm
    H.c.}\Big)\nonumber \\
  &&
\end{eqnarray}
Comparison with the effective Hamiltonians in the
dimer+hole+fermion model (for large $W$) above shows that one can
obtain the same form at the quadratic level by taking a saddle
point with $iA_{i0} = v\eta_i$, $Q_{ij},\overline{Q}_{ij}$
non-zero only when $i,j$ are on the same sublattice, and
$\Delta_{ij},\overline{\Delta}_{ij}$ non-zero only when $i,j$ are
on opposite sublattices.  For the $\pi$-flux case,
$Q_{ij},\overline{Q}_{ij}$ actually are non-zero only between
sites separated by an even number of lattice spacings in the $x$
and $y$ directions (corresponding to the vanishing of the $t'$
couplings with this background, and hence to a four-sublattice
structure of the bosons).  Phase fluctuations of these matrix
fields become the spatial components of the gauge fields of the
dimer+hole+fermion model.

Next, consider the SU(2) slave particle
formulation.\cite{su2,su2long} In this case, one represents the
electrons by an SU(2) doublet of bosons, $b_{i\sigma}$, and a
spin-$1/2$ fermion $d_{is}$, with the electron represented as
\begin{equation}
  \label{eq:elsu2}
  c_{is} = \frac{1}{\sqrt{2}}\left( b_{i+}^\dagger d_{is} +
    b_{i-}^\dagger \epsilon_{ss'} d_{is'}^\dagger \right).
\end{equation}
Up to a sign change in $b_{i-}$ (due to our trivially different
conventions), this is the same form as Eq.~(\ref{eq:electron2}).
The slave particles are constrained by $\psi_{is}^\dagger
\frac{\vec\tau}{2} \psi_{is}^{\vphantom\dagger} + b_i^\dagger
\vec\tau b_i^{\vphantom\dagger}=0$, where the doublet
\begin{equation}
  \label{eq:psidoub}
  \psi_{is} = \left(
    \begin{array}{c}
      d_{is} \\ \epsilon_{ss'} d_{is'}^\dagger
    \end{array} \right),
\end{equation}
and $\vec\tau$ is the vector of Pauli matrices acting in the
doublet space.  These constraints lead to three Lagrange
multipliers $A_{i0}^p$ ($p=1,2,3$):
\begin{equation}
  \label{eq:llma}
  {\cal L}_0^{SU(2)} = -\sum_{ip} i \vec{A}_{i0} \cdot \left( \psi_{is}^\dagger
  \frac{\vec\tau}{2} \psi_{is}^{\vphantom\dagger} + b_i^\dagger \vec\tau
  b_i^{\vphantom\dagger} \right)\; .
\end{equation}
Of particular interest is the $p=3$ component of this constraint
which takes the form
\begin{equation}
b^{\dagger}_{i+}b^{\vphantom\dagger}_{i+}-b^{\dagger}_{i-}b^{\vphantom\dagger}_{i-}
+ \sum_s d^{\dagger}_{is} d^{\vphantom\dagger}_{is}= 1.
\label{constraint4}
\end{equation}
Comparing Eq.~(\ref{constraint4}) with our
Eqs.~(\ref{constraint2}), (\ref{constraint3}), we see that they
differ in that Eqs.~(\ref{constraint2}), (\ref{constraint3})
contain an additional fluctuating electric field contribution.
This is a reflection of the different manner in which the gauge
fluctuations are realized on the lattice, rather than a
fundamental distinction. If we begin with a theory with a
constraint as in Eq.~(\ref{constraint4}), and imagine integrating
out some high energy degrees of freedom, then the gauge invariance
permits the electric field contribution in
Eqs.~(\ref{constraint2}), (\ref{constraint3}) to appear in the
renormalized theory; thus, the main difference is that in our
dimer model approach this contribution in already included in the
bare theory.

As we will see in Section~\ref{sec:dimerdual}, in our duality
analysis the electric field term is a large contribution to the
l.h.s. of Eq.~(\ref{constraint2}), which dominates that of the $f$
fermions. In contrast, in the mean-field saddle points of
Refs.~\onlinecite{su2,su2long}, the dominant contribution to
Eq.~(\ref{constraint4}) arises from the fermion $d^{\dagger}_{is}
d^{\vphantom\dagger}_{is}$ term, with the $d$ fermions occupying a
half-filled band (the $d$ fermions also contribute a large term to
the duality analysis of Section~\ref{sec:dimerdual} in
Eq.~(\ref{constraint3}), but not in a manifestly translationally
invariant manner.\cite{fdfermion}). We will comment further on the
physical distinction between these two approaches in
Section~\ref{sec:dimerdual}. Decoupling the hopping and Heisenberg
interactions in the Hamiltonian gives quadratic terms for the
slave particles of the form:
\begin{eqnarray}
  \label{eq:ldecsu2}
  {\cal L}_1^{SU(2)} & = & \sum_{ij} c_1 \psi_{is}^\dagger U_{ij}
  \psi_{js}^{\vphantom\dagger} + c_2 b_i^\dagger U_{ij}
  b_j^{\vphantom\dagger},
\end{eqnarray}
with
\begin{equation}
  \label{eq:Udef}
  U_{ij} = \left(
    \begin{array}{cc}
      -\chi_{ij}^* & \Delta_{ij} \\
      \Delta_{ij}^* & \chi_{ij}
    \end{array} \right).
\end{equation}
Our dimer+hole+fermion model described by
$\mathcal{H}_{qd}+\mathcal{H}_f$ in Eqs.~(\ref{hqd}),
(\ref{eq:Hfd}) can be described in the above SU(2) formulation by
taking a saddle point with $-iA_{i0}^3=v\eta_i$. In particular,
our physical interpretation requires that we take the saddle point
value $-v\sim W$ large so that the $b_{i\sigma}$ holes reside
primarily on their respective sublattices. In contrast, in
previous work \cite{su2,su2long} on the SU(2) gauge theory
superficially different saddle points were chosen: they worked
with a spatially uniform value of $-iA_{i0}^3$ so that only one
species of the bosons was preferentially occupied in the ground
state. However, saddle points in these two classes can be mapped
onto each other by a SU(2) gauge transformation generated by the
unitary matrix $i \tau^1$ acting only on the $\eta_i=-1$
sublattice. This gauge transformation maps our staggered
$-iA_{i0}^3$ saddle point to a spatially uniform saddle point. The
same gauge transformation acting on ${\cal
  L}_1^{SU(2)}$ interchanges the fermion hopping, $\chi_{ij}$, and
pairing, $\Delta_{ij}$, terms. So to obtain a saddle point of ${\cal
  L}_1^{SU(2)}$ consistent with our Eq.~(\ref{eq:Hfd}) we need to take
the $\Delta_{ij}$ of a form consistent with the $t_{i\alpha}$ in
Eq.~(\ref{eq:Hfd}).  The $\chi_{ij}$ will be proportional to the
$\tilde{t}$ times the square of the $b_i$ condensate.

When the model is undoped and the holons are gapped, $\langle
b_i\rangle=0$, such states fit into Wen's PSG classification
scheme for symmetric spin liquid states on the square lattice.
What are called the zero and $\pi$ flux states in this paper
correspond, in the notation of Ref.~\onlinecite{wenpsg}, to the
states U1C$n00x$, and U1C$n0n1$, respectively (see Eqs~(122) and
(135) of Ref.~\onlinecite{wenpsg}). These states are distinct in
their PSGs from the ``staggered flux'' and ``pi flux'' states of
Refs.~\onlinecite{su2,su2long}, which are denoted U1C$n01n$ and
SU2B$n0$, respectively in Ref.~\onlinecite{wenpsg}. Note that in
Wen's discussion, the spin liquids U1C$n00x$, and U1C$n0n1$ obey
the constraint in Eq.~(\ref{constraint4}); consequently the
fermion ``chemical potential'' $v$ in Eq.~(\ref{eq:Hf}) or
Eq.~(\ref{eq:Hfd}) is not an arbitrary parameter and must take a
value that ensures that Eq.~(\ref{constraint4}) is obeyed. In our
discussion, we have the additional fluctuating electric field
contribution in Eq.~(\ref{constraint3}), and we treat
Eqns.~(\ref{eq:Hf}) and (\ref{eq:Hfd}) as generic effective
Hamiltonians; large positive values of $v$, which lead to a large
spin gap and a small density of $f$ fermions, are also allowed.

\subsection{Superconducting state}
\label{sec:superc-state}

To obtain superconducting states of the doped (or undoped) dimer model,
one clearly must condense some charged particle.  In fact, a
conventional superconducting state requires a condensation of both
flavors of bosons, $\langle b_{i\sigma}\rangle \neq 0$ for both
$\sigma=\pm$.  Condensation of only one of the flavors breaks only the
combination of physical U(1) charge symmetry (since each holon carries
charge $+e$) and gauge U(1) symmetry.  The orthogonal linear combination
is unbroken in that case.  Explicitly, one can see this from
Eq.~(\ref{eq:electron2}) and Eq.~(\ref{eq:Hfd}).  When only one boson is
condensed, say $\langle b_{i+} \rangle \neq 0$, then at mean-field
level, the electron and spinon become equivalent, in this case $c_{is}
\sim \langle b_{i+} \rangle d_{is}$.  Replacing $d_{is}$ by $c_{is}$ in
Eq.~(\ref{eq:Hfd}) and dropping the $\tilde{t}$ term which vanishes
since one of the two bosons is uncondensed, one sees that one has no
anomalous electron pairing terms.  One might also imagine condensing not
the individual holon fields but instead just the quadratic forms
$\langle b_{i+}^\dagger b_{j-}^{\vphantom\dagger}\rangle \neq 0$ and
$\langle b_{i+}^\dagger b_{j-}^{\dagger}\rangle \neq 0$.  This will
certainly break the charge U(1) symmetry, and indeed follows if we
condensed both holons individually.  However, if the individual holons
remain uncondensed, there is a residual unbroken $Z_2$ gauge symmetry
under $b_{i\sigma}\rightarrow z_i b_{i\sigma}$, with $z_i = \pm 1$.
Thus such a state is an anomalous superconductor (SC$^*$), which we do
not wish to consider here.

It is simplest to think about the ``conventional'' superconducting
state in the $b_i$ and
$f_{is}$ variables.  Condensation of both flavors of holon then simply
means that $\langle b_i \rangle$ is non-zero on both sublattices.  Then,
from Eq.~(\ref{eq:electrondef}), we can regard the $f_{is}$ variables as
essentially electrons.  Formally, if the superconducting state does not
break translational symmetry, the holon condensate will have uniform
amplitude,
\begin{equation}
  \label{eq:holonc}
  \langle b_i \rangle = |b|e^{-i\phi_i},
\end{equation}
so we shift $f_{is} \rightarrow e^{-i\phi_i} f_{is}$, after which
$c_{is} \sim |b| f_{is}$.  This gives
\begin{eqnarray}
  \label{eq:Hf1}
  {\cal H}_{f}^{(1)} & \rightarrow & v\sum_{i} f_{is}^\dagger
  f_{is}^{\vphantom\dagger} \\ && +\sum_{i\alpha} \delta_\alpha e^{i
    (\phi_i + \phi_{i+\hat{e}_\alpha})} f_{is}^\dagger \epsilon_{ss'}
  f_{i+{\hat{x}_\alpha},s'}^\dagger + {\rm H.c.}, \nonumber
\end{eqnarray}
where we have dropped the gauge field which is gapped by the Higgs
mechanism in this phase.  The rescaled fermion $f_{is}$ can now be
regarded as the electron quasiparticle operator, so that the
pairing state can be read directly off from Eq.~(\ref{eq:Hf1}). We
see that the pair amplitude obeys
\begin{equation}
  \label{eq:pairamp}
  \Psi_{i\alpha}^\ast \sim \delta_\alpha e^{i
    (\phi_i + \phi_{i+\hat{e}_\alpha})}.
\end{equation}
This is in full agreement with the symmetry discussion in
Sec.~\ref{sec:proj-real-symm}.

Note that for the case of a zero flux background, the pairing can
be of $d_{x^2-y^2}$ symmetry.  This occurs e.g. for a uniform
holon condensate $\phi_i=0$ and $\delta_x=-\delta_y=1$
($e^{2i\chi(R)}=-1$), or for $\phi_i = \pi i_x$ and
$\delta_x=\delta_y=1$ ($e^{2i\chi(R)}=1$).  Nevertheless, the
spectrum of ${\cal H}_f$ is fully gapped in the limit relevant to
the doped dimer model, since there $v \gg \delta_\alpha$.  This
corresponds to a ``molecular'', ``BEC'', or ``strong pairing''
d-wave phase, where it is well-known that gapless nodal
excitations are absent.

It is amusing to consider the energetics behind the pairing
symmetry for one example.  Let us take for simplicity zero
background flux, and choose the PSG with $\delta_x=\delta_y$,
$e^{2i\chi(R)}=1$.  To apply Eq.~(\ref{eq:pairamp}), we need to
understand the pattern of holon condensation.  For simplicity, let
us ignore explicit (non-gauge mediated) holon-spinon interactions,
and gauge fluctuations.  Then we may consider the holons on their
own, described by a hopping model. We imagine lowering the
chemical potential for holons until they condense, the condensate
being built out of the lowest energy modes. Consider a single
holon excitation.  For simplicity, we consider the large $W$
limit, so that the flavor $1$ holons prefer to sit on the A
sublattice and the other flavor sits on the B sublattice.  For
$W=\infty$, the lowest one-holon states are degenerate, consisting
of any superposition of one holon of type 1 on the A sublattice,
or any superposition of one holon of type 2 on the B sublattice.
One can then apply degenerate perturbation theory to the holon
hopping terms, to arrive at an effective model of the type 1 holon
hopping on the A checkerboard sublattice and the type 2 holon
hopping on the B checkerboard sublattice.  One expect that,
allowing for generalized hoppings in the original model, this
hopping Hamiltonian takes the form
\begin{equation}
  \label{eq:hopguess}
  {\cal H}_{1h} = t_{\rm eff} \left.\sum_{\langle ij\rangle}\right.^\prime
  b_{i\sigma}^\dagger b_{j\sigma}^{\vphantom\dagger} +\cdots,
\end{equation}
where the prime on the sum indicates that the sum is over nearest
neighbors of the A(B) sublattice for $\sigma=1$ ($\sigma=2$).  The
ellipses indicate other further neighbor terms that will generally
be present.  The single-holon eigenstates of
Eq.~(\ref{eq:hopguess}) are plane waves, with energy $E_\sigma(k)=
2t_{\rm eff}(\cos (k_x+k_y) + \cos(k_x-k_y))$ (momenta are defined
for the original square lattice).  For $t_{\rm eff}<0$, the
minimum energy states have either $k_x=k_y=0,\pi$.  These two are
actually equivalent, and represent states with constant $b_i$ on
the A or B sublattice.  For $t_{\rm eff}>0$, the minimum energy
states have $k_x=0,k_y=\pi$ or $k_x=\pi,k_x=0$, which are also
equivalent and represent states with $b_i$ alternating in sign on
nearest-neighbor site of the A or B sublattice, e.g. with
$b_{i+}=1$ on sites with $x_i,y_i$ both even and $b_{i-}=-1$ on
sites with $x_i, y_i$ both odd.  At the one particle level, one
expects then the condensate which forms on doping to be a linear
superposition of these two flavors,
\begin{equation}
  \label{eq:bsup}
  \langle b_i \rangle = \psi_1 b^{min}_{i+} + \psi_2 b^{min}_{i-},
\end{equation}
where $b_{i\sigma}^{min}$ is the minimum energy single-holon
wavefunction with support only on the $\sigma$ sublattice, either
constant or staggered on this sublattice depending upon the sign
of $t_{\rm eff}$.  The nature of the superposition that occurs --
i.e. the coefficients $\psi_1,\psi_2$ -- will be determined by
holon-holon interactions.  At a mean-field level, we can simply
construct a ``Landau'' effective action for $\psi_1,\psi_2$. Since
both flavors of bosons are separately conserved, it should have
$U(1)\times U(1)$ symmetry under rotations of each field. Further,
it should be symmetric under interchange of the two flavors, by
translation invariance, which interchanges the two sublattices.
One expects therefore an effective (Euclidean) Lagrangian of the
form
\begin{equation}
  \label{eq:Leffpsis}
  {\cal L}_{\rm eff} = - r (|\psi_1|^2 + |\psi_2|^2) + u
  (|\psi_1|^2+|\psi_2|^2)^2 - v |\psi_1|^2 |\psi_2|^2.
\end{equation}
The nature of the condensate depends upon the sign of $v$.  For
$v>0$, one obtains $|\psi_1|=|\psi_2|$, while for $v<0$, one or
the other amplitude vanishes.  Since the latter case clearly
breaks translational symmetry, we focus on the former.  Then for
$t_{\rm eff}<0$, the phases $\phi_i$ are constant on all sites of
one sublattice, which for $t_{\rm eff}>0$, the $\phi_i$ alternate
between $0$ and $\pi$ on the neighboring (diagonal second
neighbors on the original square lattice) sites of a given
sublattice.

Now let us consider the pairing symmetry.  As we assumed
$\delta_y=\delta_x$, we have from Eq.~(\ref{eq:pairamp})
\begin{equation}
  \label{eq:ratio}
  \frac{\Psi_{iy}}{\Psi_{ix}} =
  e^{i(\phi_{i+\hat{e}_x}-\phi_{i+\hat{e}_y})}.
\end{equation}
Since $i+\hat{e}_x$, $i+\hat{e}_y$ are nearest neighbor sites on
the same sublattice, one finds
\begin{equation}
  \label{eq:ratio2}
  \frac{\Psi_{iy}}{\Psi_{ix}} =
  \left\{
    \begin{array}{rl}
      1 & \mbox{for $t_{\rm eff}<0$}, \\
      -1 & \mbox{for $t_{\rm eff}>0$}
    \end{array} \right.
\end{equation}
Thus the pairing is $s$-wave or $d$-wave in these two cases,
respectively.

\section{Duality analysis of the doped dimer model}
\label{sec:dimerdual}

This section returns to an analysis the of effective theory of the
total spin $S=0$ sector presented in Section~\ref{sec:dimer}. For
this limited theory, we claim that the following analysis properly
accounts for the strong fluctuations of the compact U(1) gauge
field $\hat{\mathcal{A}}$; a corresponding, non-perturbative
treatment of gauge fluctuations has not so far been possible in
the slave particle theories.\cite{sr,su2,vs,ps,mv,su2long,wenpsg}
The $S=1/2$ fermions $f_{is}$ are neglected below,\cite{fdfermion}
and it is clear that this is surely safe as long as the $f_{is}$
remain fully gapped. We noted in Section~\ref{sec:fermions} that
the structure of our theory allows the appearance of gapless nodal
fermionic excitations in the superconducting phases, and that they
do not significantly modify the structure of the resulting
superconductor. As long as such nodal excitations are present only
for a $\delta > \delta_f$ for which the ground state is a
superconductor or a supersolid, the following analysis will apply
for the full range of $\delta$. However, it is also possible that
$\delta_f$ co-incides with the position of a
superconductor-insulator transition: such a critical point will
not be described by the theories presented below and in I.

Before embarking upon our duality analysis, we remark further on
why the neglect of the $f_{is}$ fermions is not as dangerous at it
might na\"ively seem, and why it is amply compensated by the
non-perturbative treatment of gauge fluctuations. It is useful to
refer back to our key constraint equations
Eq.~(\ref{constraint2}), (\ref{constraint3}), and the
corresponding constraint equation for the slave particle theory in
Eq.~(\ref{constraint4}). In the earlier slave particle theories,
the dominant, order unity, and translationally invariant
contribution to the l.h.s. of Eq.~(\ref{constraint4}) is provided
by the fermion density term $d^{\dagger}_{is}
d^{\vphantom\dagger}_{is}$. However, by focusing on a static gauge
field saddle point, these theories do not allow a significant
contribution to this constraint equation from the $\Delta_\alpha
\hat{E}_{i \alpha}$ term on the l.h.s. of
Eqs.~(\ref{constraint2}), (\ref{constraint3}) (such contributions
are only generated upon including fluctuations about the static
gauge saddle points). In our duality approach below (and in the
previous work \cite{rs,fradkiv,curreac} on the undoped model) a
large contribution is given by the $\Delta_\alpha \hat{E}_{i
\alpha}$ term: this is only possible because we account for strong
gauge fluctuations. The contribution of the $f$ fermionic term in
Eq.~(\ref{constraint2}), from either nodal or gapped excitations,
is small (note, however, that the contribution of the $d$ fermion
term in our Eq.~(\ref{constraint3}) is {\em not\/} small
\cite{fdfermion}), and its neglect appears safe. Physically
speaking, the $\Delta_\alpha \hat{E}_{i \alpha}$ term counts the
valence bonds on the links emerging from a site, and so our
approach here assumes that most of the electrons appear in singlet
pairs, rather than individually in single particle states.

We proceed with a duality mapping of $\mathcal{H}_{qd}$ as in
Ref.~\onlinecite{curreac} and similar to that described in I. The
cosine term in $\mathcal{H}_t$ is written in the Villain form as
\begin{eqnarray}
&& \exp\left(t \Delta \tau \cos \left( \Delta_{\alpha}
\hat{\phi}_{i\sigma} + \sigma \hat{\mathcal{A}}_{i\alpha} \right)\right) \label{f4} \\
&& \!\!\!\!\!\!\!\!\!\!\!\!\! \rightarrow
\sum_{\{J_{i\sigma\alpha}\}} \exp \left( -
\frac{J_{i\sigma\alpha}^2}{2 t \Delta \tau} + i J_{i \sigma\alpha}
\Delta_{\alpha} \hat{\phi}_{i\sigma} + i \sigma J_{i \sigma
\alpha} \hat{\mathcal{A}}_{i \alpha} \right), \nonumber
\end{eqnarray}
while that in $\mathcal{H}_d$ is written as
\begin{eqnarray}
&& \exp\left(-J \Delta \tau \cos \left(\epsilon_{\alpha\beta}
\Delta_{\alpha}
\hat{\mathcal{A}}_{i \beta}\right)\right) \rightarrow \nonumber \\
&&\!\!\!\!\!\!\!\!\!\! \sum_{\{B_a\}} \exp \left( - \frac{B_a^2}{2
J \Delta \tau} + i B_a \epsilon_{\alpha\beta} \Delta_{\alpha}
\hat{\mathcal{A}}_{i \beta} + i \pi B_a  \right). \label{f4a}
\end{eqnarray}
Here $B_a$ is an integer-valued dual magnetic field on the sites
of the dual lattice. Then we integrate over the ${\phi}_{i\sigma}$
and $\mathcal{A}_{i\alpha}$ and obtain the dual partition function
\begin{eqnarray}
&& \mathcal{Z}_{qd} = \sum_{\{J_{i\sigma\mu}, F_{a\mu}\}} \exp
\Biggl(- \frac{g}{2} \sum_{a}  F_{a \mu}^2 -  i \pi \sum_a F_{a\tau}  \nonumber \\
&&~~~~~- \frac{1}{2 e^2} \sum_{i\sigma} \left[ J_{i \sigma \mu} -
(H+ \sigma \eta_i W ) \delta_{\mu\tau} \right]^2 \nonumber
\\
&&~~~~~-\Delta \tau \sum_{i\neq j,\sigma} \Lambda_{ij}
J_{i\sigma\tau} J_{j\sigma\tau} \Biggr) \prod_{i\sigma} \delta
\left( \Delta_\mu J_{i\sigma\mu} \right) \nonumber
\\ &&~~~\times \prod_{a\mu} \delta \left( \epsilon_{\mu\nu\lambda}
\Delta_\nu F_{a \lambda} + \sum_\sigma \sigma J_{i\sigma\mu} -
\eta_i \delta_{\mu\tau} \right) \label{zdd}
\end{eqnarray}
where $J_{i\sigma\mu} = ( n_{\sigma}, J_{i\sigma x}, J_{i\sigma
y})$ represent the integer valued spacetime currents of the two
species of holes, $F_{a\mu} = (- B_a,-E_{iy}, E_{ix}, )$ is the
integer valued gauge flux, $g = 1/(J \Delta \tau) = V \Delta
\tau$, and as in I, $e^2 = t \Delta \tau = 1/U \Delta \tau$. The
$i \pi F_{a \tau}$ term will be innocuous in the present
subsection, and we will return to a consideration of its effects
in Section~\ref{sec:dwave} when we examine the signature of the
electronic pairing.

\begin{figure}
\centering
\includegraphics[width=3in]{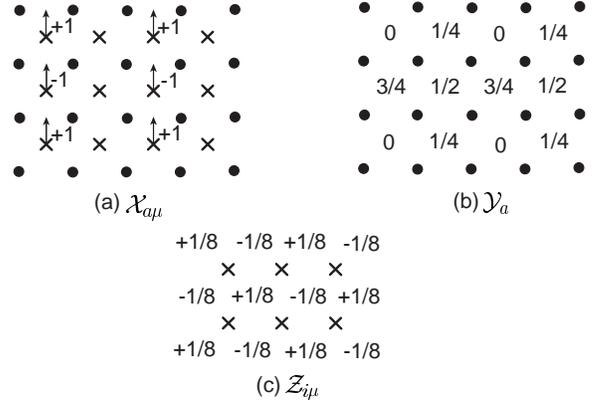}
\caption{The values of the only non-zero components of the fixed
field $\mathcal{X}_{a\mu}$, $\mathcal{Y}_a$, and
$\mathcal{Z}_{i\mu}$. The circles (crosses) are the sites of the
direct (dual) lattice. In (c), only the $\mu=\tau$ component of
$\mathcal{Z}_{i\mu}$ is non-zero and its values are shown.}
\label{offsets}
\end{figure}

We solve the constraints in Eq.~(\ref{zdd}) by parameterizing
\begin{eqnarray}
J_{i \sigma \mu} &=& \epsilon_{\mu\nu\lambda} \Delta_\nu b_{a
\sigma \lambda} \nonumber \\
F_{a \mu} &=& \Delta_\mu N_a + \mathcal{X}_{a\mu} - \sum_{\sigma}
\sigma b_{a \sigma \mu} \label{solvecons}
\end{eqnarray}
The integer-valued fixed field $\mathcal{X}_{a\mu}$ is defined as
in Ref.~\onlinecite{curreac} and is shown in Fig~\ref{offsets}; it
obeys the constraint
\begin{equation}
\epsilon_{\mu\nu\lambda} \Delta_{\nu} \mathcal{X}_{a \lambda} =
\eta_i \delta_{\mu\tau} \label{eq:curlX}
\end{equation}
The degrees of freedom are integers $b_{a \sigma \mu}$ on the
links of the dual lattice and integers $N_a$ on the sites of the
dual lattice.

Now we promote these integer-valued fields to real fields by
introducing a vortex fugacity $y_v$ and a monopole fugacity $y_m$.
Then
\begin{eqnarray}
&& \mathcal{Z}_{qd} = \prod_{a \sigma} \int d b_{a \sigma \mu}
\int d \vartheta_{a \sigma} \int d N_{a} \exp \Biggl[ \nonumber
\\
&&~~~~~~- \frac{g}{2} \sum_a \left(\Delta_\mu N_a +
\mathcal{X}_{a\mu} - \sum_{\sigma} \sigma b_{a \sigma \mu}
\right)^2 \nonumber \\
&&~~~~~~- i \pi \sum_a \left(\Delta_\tau N_a + \mathcal{X}_{a\tau}
- \sum_{\sigma} \sigma b_{a \sigma \tau}
\right) \nonumber \\
&&~~~~~~- \frac{1}{2 e^2} \sum_{i\sigma} \left[
\epsilon_{\mu\nu\lambda} \Delta_\nu b_{a \sigma \lambda} -
(H+ \sigma \eta_i W ) \delta_{\mu\tau} \right]^2 \nonumber \\
\nonumber
\\
&&~~~~~~-\Delta \tau \sum_{a\neq a',\sigma} \Lambda_{ij}
\epsilon_{\tau\nu\lambda} \Delta_{\nu} b_{a\sigma \lambda}
\epsilon_{\tau\rho\varrho} \Delta_{\rho}
b_{a'\sigma \varrho}\nonumber \\
&&~~~~~~+y_v \sum_{a\sigma} \cos ( \Delta_{\mu} \vartheta_{a
\sigma} - 2 \pi b_{a \sigma \mu} ) \nonumber \\
&&~~~~~~+ y_m \sum_a \cos \left( 2 \pi N_a - \sum_{\sigma} \sigma
\vartheta_{a \sigma} \right) \Biggr] \label{zde}
\end{eqnarray}
This is a theory of 2 vortex fields, $\psi_{+} = e^{i
\vartheta_{+}}$ and $\psi_{-}= e^{i \vartheta_{-}}$ which are
coupled to 2 non-compact U(1) gauge fields $b_{+\mu}$ and
$b_{-\mu}$. There is also a Higgs field $N_a$ which breaks the
staggered component of the U(1) gauge symmetries, as will become
more explicit below.

We will henceforth not carry through the off-site $\Lambda_{ij}$
terms explicitly in our analysis. As in I, we will assume that
their effects can be absorbed into a renormalization of the value
of $e^2$. This is certainly the case for short-range
$\Lambda_{ij}$, while for long-range Coulomb interactions, $e^2$
will acquire momentum dependence which was noted in I. Our
numerical results below will be restricted to the short-range
case, but we do not expect significant modifications for the
long-range case.

To allow us to make direct contact with previous work on the
undoped dimer model, we use the parameterization of
$\mathcal{X}_{a\mu}$ in Ref.~\onlinecite{curreac} in terms of
curl-free and divergence-free fields (shown in Fig~\ref{offsets})
\begin{equation}
\mathcal{X}_{a \mu} = \Delta_{\mu} \mathcal{Y}_a + \epsilon_{\mu
\nu \lambda} \Delta_{\nu} \mathcal{Z}_{i \lambda}. \label{f7}
\end{equation}
Inserting this in Eq.~(\ref{zde}) and shifting $N_a \rightarrow
N_a - \mathcal{Y}_a$ we obtain
\begin{eqnarray}
&& \mathcal{Z}_{qd} = \prod_{a \sigma} \int d b_{a \sigma \mu}
\int d \vartheta_{a \sigma} \int d N_{a} \exp \Biggl[ \nonumber
\\
&&~~~~~~- \frac{g}{2} \sum_a \left(\Delta_\mu N_a +
\mathcal{E}_{a\mu} - \sum_{\sigma} \sigma b_{a \sigma \mu}
\right)^2 \nonumber \\
&&~~~~~~- i \pi \sum_a \left(\Delta_\tau N_a + \mathcal{E}_{a\tau}
- \sum_{\sigma} \sigma b_{a \sigma \tau}
\right) \nonumber \\
&&~~~~~~- \frac{1}{2 e^2} \sum_{i\sigma} \left[
\epsilon_{\mu\nu\lambda} \Delta_\nu b_{a \sigma \lambda} -
(H+ \sigma \eta_i W ) \delta_{\mu\tau} \right]^2 \nonumber \\
&&~~~~~~+y_v \sum_{a\sigma} \cos ( \Delta_{\mu} \vartheta_{a
\sigma} - 2 \pi b_{a \sigma \mu} ) \nonumber \\
&&~+ y_m \sum_a \cos \left( 2 \pi N_a - \sum_{\sigma} \sigma
\vartheta_{a \sigma} - 2\pi \mathcal{Y}_a \right) \Biggr],
\label{zde1}
\end{eqnarray} where $\mathcal{E}_{a\mu} =
\epsilon_{\mu \nu \lambda} \Delta_{\nu} \mathcal{Z}_{i \lambda}$
is given by
\begin{equation}
\mathcal{E}_{a\mu} \equiv \frac{(-1)^{a_x+a_y}}{4} (0,1,-1).
\end{equation}
It is convenient to introduce uniform ($A_\mu$) and `staggered'
($C_\mu$) gauge fields defined by
\begin{eqnarray}
b_{a+\mu} &=& A_{a\mu} + C_{a\mu} \nonumber \\
b_{a-\mu} &=& A_{a\mu} - C_{a\mu}\;. \label{bAC}
\end{eqnarray}
The term proportional to $g$ in Eq.~(\ref{zde1}) effectively
breaks the `staggered' gauge symmetry associated with $C_\mu$
under which $\psi_{\pm}$ have opposite gauge charges. The field
$N_{a}$ is the phase of the Higgs field which breaks this
symmetry, and as is conventional, we use a gauge transformation to
set this field equal to 0. Also, we transform from the hard-spin
to soft-spin variables to obtain the dual action in its final form
\begin{eqnarray}
&& \mathcal{Z}_{qd} = \prod_{a \mu} \int d A_{a \mu} \int d
C_{a\mu} \prod_{a \sigma} \int d \psi_{a \sigma} \exp \Biggl[
\nonumber \\
&&~~~~~~- \frac{1}{e^2} \sum_{a} \left[ \epsilon_{\mu\nu\lambda}
\Delta_\nu A_{a \lambda} - H
\delta_{\mu\tau} \right]^2 \nonumber \\
&&~~~~~~- \frac{1}{e^2} \sum_{i} \left[ \epsilon_{\mu\nu\lambda}
\Delta_\nu C_{a \lambda} - \eta_i W \delta_{\mu\tau} \right]^2
\nonumber \\
&&~~~~~~- \frac{g}{2} \sum_a \left( 2 C_{a\mu} -
\mathcal{E}_{a\mu} \right)^2 + 2 i \pi \sum_a C_{a \tau} \nonumber
\\
&& -\sum_{a\sigma} \left[ s |\psi_{a\sigma} |^2 + \frac{u}{2}
|\psi_{a\sigma}|^4 \right] -v \sum_a |\psi_{a+}|^2 |\psi_{a-}|^2
\nonumber \\
&&~~~+ \frac{y_v}{2} \sum_{a \sigma \mu} \left[
\psi_{a+\mu,\sigma}^{\ast} e^{2 \pi i (A_{a \mu}+\sigma C_{a\mu})}
\psi_{a\sigma} + \mbox{c.c.} \right]
\nonumber \\
&&~~~+\frac{y_m}{2} \sum_a \left[ \psi_{a-}^{\ast} \psi_{a+} e^{2
\pi i \mathcal{Y}_a} + \mbox{c.c.} \right] \Biggr]. \label{zde2}
\end{eqnarray}
The similarity between $\mathcal{Z}_{qd}$ above, and the dual theory of
bosons on a square lattice in I should now be evident.  The latter
theory had a single vortex species, $\psi_a$, coupled to a single
non-compact U(1) gauge field $A_{a \mu}$. Here in Eq.~(\ref{zde2}) we
have two vortex species, $\psi_{a+}$ and $\psi_{a-}$, but they are
coupled together by the monopole fugacity term proportional to $y_m$,
effectively reducing the theory to that of a single vortex field. The
present partition function also has 2 non-compact U(1) gauge fields,
$A_{a \mu}$ and $C_{a \mu}$, but the $C_{a \mu}$ field has acquired a
mass from the Higgs phenomenon, as is clear from the term proportional
to $g$.  This massive gauge field also has a complex Berry phase term
(with $2 i \pi$ in Eq.~(\ref{zde2})), which is unimportant here because
the $C_{a\mu}$ fluctuations are quenched.  The physical meaning of this
quenching was already discussed at the end of Sec.~\ref{sec:dimer}: in
this representation, the regime of interest where the dimer resonance
move has relatively small amplitude corresponds to large $g$, and the
$\pi$ gauge flux favored by this term (responsible for the complex Berry
phase term just mentioned) is expelled.

Given the close similarity between $\mathcal{Z}_d$ in I and
$\mathcal{Z}_{qd}$ in Eq.~(\ref{zde2}), it is evident that low
energy fluctuations in the vicinity of transitions between
superfluid, supersolid, and insulating phases of the quantum dimer
model are described by the {\em same\/} continuum quantum field
theories as those in I. The most important fact determining the
character of these field theories is the particular projective
representation of the square lattice space group (PSG) that is
realized by the saddle point under consideration. In particular,
all we need is the value of the effective dual `magnetic' flux
$f$, and the corresponding value of the unimodular complex number
$\omega$ in I. Determining this requires a careful symmetry
analysis of the appropriate saddle point of Eq.~(\ref{zde2}), and
this is carried out in the following subsection.

Before turning to this symmetry analysis, it is useful to make
explicit contact between the dual partition function in
Eq.~(\ref{zde2}), and previous dual representations
\cite{rs,curreac} of the {\em undoped} dimer model. The hole
density vanishes in this limit, and so the $A_{a \mu}$ flux is
zero; this is achieved by setting $H=0$ in Eq.~(\ref{zde2}).
Furthermore, the insulating behavior requires a strong condensate
of the $\psi_{a \pm}$ fields, and we can focus on the phase
fluctuations of the condensate by writing $\psi_{a \pm} = e^{i
\chi_{a \pm}}$. We also set the massive field $C_{a\mu} = 0$. Then
the action $\mathcal{Z}_{qd}$ maps at low energies to
\begin{eqnarray}
&& \mathcal{Z}_{qd} = \prod_{a \mu} \int d A_{a \mu} \prod_{a
\sigma} \int d \chi_{a \sigma} \exp \Biggl[
\nonumber \\
&&~~~~~~- \frac{1}{e^2} \sum_{a} \left[ \epsilon_{\mu\nu\lambda}
\Delta_\nu A_{a \lambda} \right]^2 \nonumber \\
&&~~~~~~- \frac{y_v}{2} \sum_{a \mu}  \left( \frac{\Delta_\mu
\chi_{a+} + \Delta_\mu \chi_{a-}}{2} + 2 \pi A_{a \mu} \right)^2
 \nonumber \\
&&~~~~~~- \frac{y_v}{8} \sum_{a \mu} \left( \Delta_\mu \chi_{a+} -
\Delta_\mu \chi_{a-} \right)^2
\nonumber \\
&&~~~~~~~~~~+y_m \sum_a \cos\left( \chi_{a+} - \chi_{a-} + 2 \pi
\mathcal{Y}_a \right) \Biggr]. \label{zde3}
\end{eqnarray}
The first two terms show that the $A_{\mu}$ gauge field is Higgsed
by $\chi_{a+} + \chi_{a-}$, and so is innocuous. The last two
terms constitute a sine-Gordon model for $\chi_{a+} - \chi_{a-}$,
with on-site offset $\mathcal{Y}_a$: this is precisely the dual
representation of the undoped dimer model found earlier
\cite{rs,curreac}.

\subsection{Symmetries}
\label{sec:dimersym}

As in I, we will analyze the symmetry properties of the dual dimer
model theory in Eq.~(\ref{zde2}) at a commensurate density of
holes appropriate to a proximate Mott insulator. We assume this
has the rational value $\delta_{I}$ which we parameterize as
\begin{equation}
\frac{\delta_I}{2} = \frac{p}{q},
\end{equation}
where $p$ and $q$ are relatively prime integers. In general (as in
I), the density of holes, $\delta$, in the theory in
Eq.~(\ref{zde2}) is determined by the parameter $H$, and in the
superfluid or supersolid phases we may have $\delta \neq
\delta_{I}$.

Proceeding as in I, we set the gauge field $A_{a \mu}$ equal to a
saddle point value $\overline{A}_{a \mu}$ such that the flux is
equal to the density of each species of hole ($+$ or $-$). So we
choose $\overline{A}_{a\tau} = \overline{A}_{ax} = 0$ and
\begin{equation}
\overline{A}_{ay} = \frac{\delta_I}{2} ~a_x. \label{landau}
\end{equation}
For the staggered gauge field we have the saddle point value
$\overline{C}_{a\mu} = \frac{\lambda}{2} \mathcal{E}_{a\mu}$ with
$\lambda = 1/(1+4/e^2 g)$.

The main new subtlety here (beyond that of I) in the symmetry
analysis of Eq.~(\ref{zde2}) is the presence of the fixed
background fields $\mathcal{E}_{a\mu}$ and $\mathcal{Y}_a$. These
fields are not explicitly invariant under the square lattice space
group, and so require additional transformations of the vortex
fields upon operations of the space group. These additional
transformation will modify the needed PSG, as shown below.

Carrying out the space group operations in the presence of the
$\overline{A}_{a\mu}$, $\mathcal{E}_{a\mu}$, and $\mathcal{Y}_a$,
it can be shown that the action remains invariant under the
following field transformations, which are the analog of the
relations in Section II.B of I:
\begin{eqnarray}
T_y &:& \psi_+ (k_x, k_y)  \rightarrow \psi_- (k_x, k_y) e^{-i
k_y} e^{i \pi/4} \nonumber
\\
T_y &:& \psi_- (k_x, k_y)  \rightarrow \psi_+ (k_x, k_y ) e^{-i
k_y} e^{-i \pi/4} \nonumber
\\
T_x &:& \psi_+ (k_x, k_y)  \rightarrow \psi_- (k_x, k_y - \pi
\delta_I) e^{-i k_x} e^{-i \pi/4} \nonumber
\\
T_x &:& \psi_- (k_x, k_y)  \rightarrow \psi_+ (k_x, k_y - \pi
\delta_I) e^{- i k_x} e^{i \pi/4} \nonumber
\\
R_{\pi/2} &:& \psi_+ (k_x, k_y)  \rightarrow
\nonumber \\
&~& \frac{1}{q} \sum_{m,n=0}^{q-1}
 \psi_{-} (k_y +  \pi n \delta_I , -k_x + \pi m \delta_I ) \omega^{- m n } \nonumber \\
R_{\pi/2} &:& \psi_- (k_x, k_y)  \rightarrow \label{ktrans2} \\
&~& \frac{1}{q} 
\sum_{m,n=0}^{q-1} \psi_{+} (k_y + \pi n \delta_I , -k_x +  \pi m
\delta_I ) \omega^{- m n } \nonumber
\end{eqnarray}
To understand the degeneracy of the spectrum associated with these
transformations, it is useful to transform to a basis of states
which are eigenvectors of $T_y$. These are expressed by the fields
\begin{eqnarray}
\psi_1 &=& \frac{1}{\sqrt{2}} \left( e^{-i \pi/8} \psi_+ + e^{i
\pi/8} \psi_- \right) \nonumber \\
\psi_2 &=& \frac{1}{\sqrt{2}} \left( e^{-i \pi/8} \psi_+ - e^{i
\pi/8} \psi_- \right).
\end{eqnarray}
Then
\begin{eqnarray}
T_y &:& \psi_1 (k_x, k_y)  \rightarrow \psi_1 (k_x, k_y) e^{-i
k_y}  \nonumber
\\
T_y &:& \psi_2 (k_x, k_y)  \rightarrow - \psi_2 (k_x, k_y ) e^{-i
k_y}  \nonumber
\\
T_x &:& \psi_1 (k_x, k_y)  \rightarrow \psi_2 (k_x, k_y - \pi
\delta_I ) e^{-i k_x} e^{i \pi/2} \nonumber
\\
T_x &:& \psi_2 (k_x, k_y)  \rightarrow \psi_1 (k_x, k_y - \pi
\delta_I ) e^{- i k_x} e^{-i \pi/2} \nonumber
\\
R &:& \psi_1 (k_x, k_y)  \rightarrow \frac{1}{q} 
\sum_{m,n=0}^{q-1} \omega^{-mn} \nonumber \\
&~&~~~~\times
 \frac{[\psi_1 + i\psi_2] (k_y + \pi n \delta_I , -k_x + \pi m \delta_I )}{\sqrt 2}
 \nonumber \\
R &:& \psi_2 (k_x, k_y)  \rightarrow -\frac{1}{q} 
\sum_{m,n=0}^{q-1} \omega^{-mn}  \label{ktrans3} \\
&~&~~~~\times \frac{[\psi_2 + i\psi_1](k_y + \pi n \delta_I , -k_x
+ \pi m \delta_I )}{\sqrt 2}. \nonumber
\end{eqnarray}
Now we observe that the $T_y$ eigenvalues of the $\psi_1$ fields
at momentum $k$ are equal to those of the $\psi_2$ fields at $k+
(0,\pi)$. So these fields will mix with each other, and cannot
lead to orthogonal eigenmodes. It is convenient to introduce a new
collective field $\Psi(k)$ which equals $\psi_1$ at $k$ and
$\psi_2$ at $k+(0,\pi)$, and which is identified by the eigenvalue
$e^{-ik_y}$ of $T_y$. Then, ignoring unimportant phase factors, we
have under the action of $T_x$:
\begin{equation}
T_x : \Psi (k_x, k_y) \rightarrow \Psi (k_x , k_y + \pi - \pi
\delta_I )
\end{equation}
From this equation we see that the PSG of the present fields has
the same structure as that in Section I of I
\begin{eqnarray}
T_x T_y &=& \omega T_y T_x , \nonumber \\
\omega &\equiv& e^{2 \pi i f}
\end{eqnarray}
with the key parameter $f$ specified by
\begin{equation}
f \equiv \frac{\tilde{p}}{\tilde{q}} = \frac{(1-\delta_I)}{2} =
\frac{1}{2} - \frac{p}{q} ,\label{phieff}
\end{equation}
where $\tilde{p}$ and $\tilde{q}$ are also relatively prime
integers. Note that the value of $f$ is half the density of {\sl
electrons} {\em i.e.\/} it is the density of Cooper pairs in the
Mott insulator. So the low energy field theory of the present
dimer model is identical to that presented in I but with the PSG
determined by the density of Cooper pairs. We have also explicitly
diagonalized the spectrum of $\psi_{a \sigma}$ vortices in
$\mathcal{Z}_{qd}$ in Eq.~(\ref{zde2}) and verified that it did
indeed have the degeneracy required by such a PSG.

\subsection{Mean field theory}
\label{sec:dimermft}

We have seen above that the low energy continuum theory of the
quantum dimer model is identical to that already presented in I.
We have already presented a mean-field analysis of this continuum
theory in Section II.D of I, and so do not need to repeat it here.
Rather, we will work directly with the lattice action in
Eq.~(\ref{zde2}) and follow its mean-field phase diagram with
increasing hole density, $\delta$.

To do a mean-field analysis of our dual dimer model we consider
the free energy corresponding to Eq.~(\ref{zde2}). Rescaling the
fields $\psi_{a\sigma}$, and neglecting an overall constant, we
can choose $u=y_v=1$ and obtain the free energy (with $r \equiv s
- 2$)
\begin{align}
\mathcal{F}  & = \frac{1}{2} \sum_{a \sigma \mu} \left|
\psi_{a+\mu,\sigma} e^{-2 \pi i (A_{a \mu}+\sigma C_{a\mu})}
- \psi_{a\sigma}  \right|^2 \nonumber \\
&~ +\sum_{a\sigma} \left[ r |\psi_{a\sigma} |^2 + \frac{1}{2}
|\psi_{a\sigma}|^4 \right] +v \sum_a |\psi_{a+}|^2 |\psi_{a-}|^2 \nonumber \\
&~-\frac{y_m}{2} \sum_a \left[ \psi_{a-}^{\ast} \psi_{a+} e^{2
\pi i \mathcal{Y}_a} + \mbox{c.c.} \right]  \nonumber \\
&~+ \frac{1}{e^2} \sum_{a} \left[ \epsilon_{\mu\nu\lambda}
\Delta_\nu A_{a \lambda} - H
\delta_{\mu\tau} \right]^2 \nonumber \\
&~+\frac{1}{e^2} \sum_{a} \left[ \epsilon_{\mu\nu\lambda}
\Delta_\nu C_{a \lambda} - \eta_{i_a} W \delta_{\mu\tau} \right]^2
\nonumber \\
&~+ \frac{g}{2} \sum_a \left( 2 C_{a\mu} - \mathcal{E}_{a\mu}
\right)^2 \;. \label{eq:F_MF}
\end{align}
For simplicity we will also set $y_m$ equal to $1$, and, as
discussed below Eq.~(\ref{constraint}), we can safely choose
$W=0$. We will set $v$ equal to zero and do not discuss states
which break the $\sigma \rightarrow -\sigma$ symmetry. Such states
break the vortex-anti-vortex symmetry on average, and emerge only
for sufficiently large $v$.

We note that we have dropped the $2 i \pi \sum_a C_{a \tau} $ term
in Eq.~(\ref{zde2}), which was a consequence of the sign $J>0$ in
the dimer `resonance' term Eq.~(\ref{hd1}). The massive $C_{a
\tau}$ field has little effect on our analysis below, and it
therefore appears reasonable to ignore this term in our mean-field
theory.

This leaves us with the hole chemical potential (the `magnetic
field') $H$ (in the superfluid phase, this picks a hole density
$\delta = 2H$), the coupling constants $e^2$ and $g$, and the
control parameter $r$. For the phase transition between the
superfluid and the insulating phase to be of second order we have
to choose $e^2$ small. In our units, we take $e^2=0.04$.

We can now map out the phase diagram for zero doping. Instead of
using $r$ and $g$ as parameters, let us use $r$ and the already
introduced (below Eq.~(\ref{landau})) parameter
\begin{equation}
  \lambda \equiv \frac{1}{1+4/e^2 g} \;.
  \label{eq:lambda}
\end{equation}
Note that the range $0 < g < \infty$ gets mapped onto
$0<\lambda<1$. Also, in the superfluid phase, and close to the
second order phase transition to this phase, the saddle point
approximation for the massive field $C_{a\mu}$ becomes exact such
that $B_{\rm staggered} =  \epsilon_{\tau \mu \nu} \Delta_{\mu}
C_{a \nu}  =  \pm\lambda/2$.

\begin{figure}[tb]
\centering
\includegraphics[width=3in]{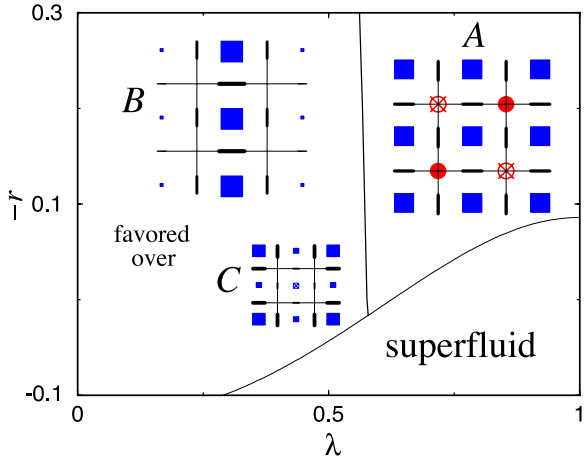}
\caption{Mean field phase diagram for the doped dimer model with
zero doping. For large $r$ our theory favors a vortex vacuum such
that we are in the superfluid phase. Decreasing $r$ we enter one
of the (condensed) insulating states. While for smaller $\lambda$
we find a columnar bond ordered state ($B$) whose free energy is
slightly lower than that of a plaquette state ($C$), for larger
values of $\lambda$ we get a CDW state ($A$). The lines indicate
the links of the direct lattice. The black bars on these links are
proportional to the spin exchange energy or the electron pairing
amplitude (see text). Similarly, the size of the (blue) boxes
depicts the ring exchange amplitudes which are located  on the
dual lattice. If this amplitude is negative we have used a (tiny)
crossed box. Finally, the state $A$ also shows a finite mean hole
density which is positive on one sublattice (red bullets) and
negative on the other (crossed circles). As discussed in the text
this CDW state is unphysical, and we exclude the corresponding
parameter values.} \label{fig:phasediagram0}
\end{figure}

As can be seen in Fig~\ref{fig:phasediagram0}, for large enough
$r$ we are always in the superfluid phase, which is characterized
by the vortex vacuum $\langle \psi_{a \mu} \rangle = 0$.
As we decrease $r$, we enter an insulating state with $\langle
\psi_{a \mu} \rangle \neq 0$ via a second order phase transition.
Depending on the value of $\lambda$, this can either be a columnar
valence bond solid (VBS) state ($B)$, or a state with site density
wave order (CDW) at wavevector $(\pi,\pi)$ ($A$). It turns out
that the plaquette state ($C$) always has a free energy slightly
larger than that of the columnar VBS state, for the parameters
chosen here. All these results are consistent with the mean field
phase diagram of our field theory for $q=2$ in I.

\begin{figure}[t]
\begin{center}
\includegraphics[width=3in]{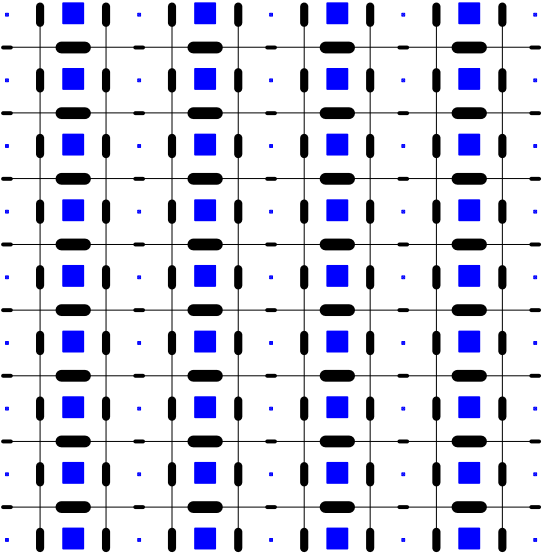}
\end{center}
\caption{Insulating state with columnar valence bond solid (VBS)
order at zero doping. This state is obtained by a mean-field
minimization of the free energy in Eq.~(\ref{eq:F_MF}) with
parameter values as discussed in the text. The significance of the
symbols on the square lattice above (and in
Figs~\ref{fig:H=0.015625}-\ref{fig:H=0.0625}) are as in
Fig~\ref{fig:phasediagram0}.} \label{fig:H=0}
\end{figure}

\begin{figure}
\begin{center}
\includegraphics[width=3in]{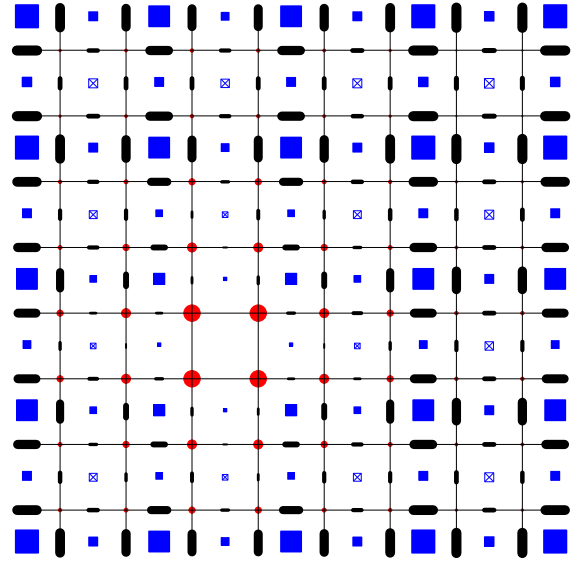}
\end{center}
\caption{An insulating state with a $8 \times 8$ unit cell for
doping $\delta=1/32$. The parameters are the same as for the
$\delta=0$ state in Fig~\protect\ref{fig:H=0}, apart from the
change in $H$, the hole chemical potential. This state is
invariant under rotations by $90$ degrees around the plaquette
with $\langle \psi_{a \sigma} \rangle = 0$.
 The mean hole density indicated by the red bullets is concentrated near this plaquette.}
\label{fig:H=0.015625}
\end{figure}

\begin{figure}[t]
\begin{center}
\includegraphics[width=3in]{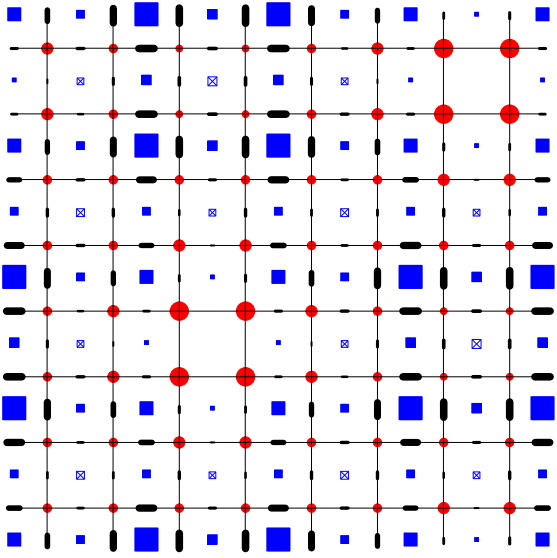}
\end{center}
\caption{Insulating state evolving out of state in
Fig~\ref{fig:H=0.015625} for doping $\delta=1/16$ (other
parameters as in Fig~\ref{fig:H=0}). As for the state depicted in
Fig~\ref{fig:H=0.015625} this state is symmetric under rotations
by $90$ degrees. However, the unit cell is now $4 \times 8$.}
\label{fig:H=0.03125}
\end{figure}

\begin{figure}
\begin{center}
\includegraphics[width=3in]{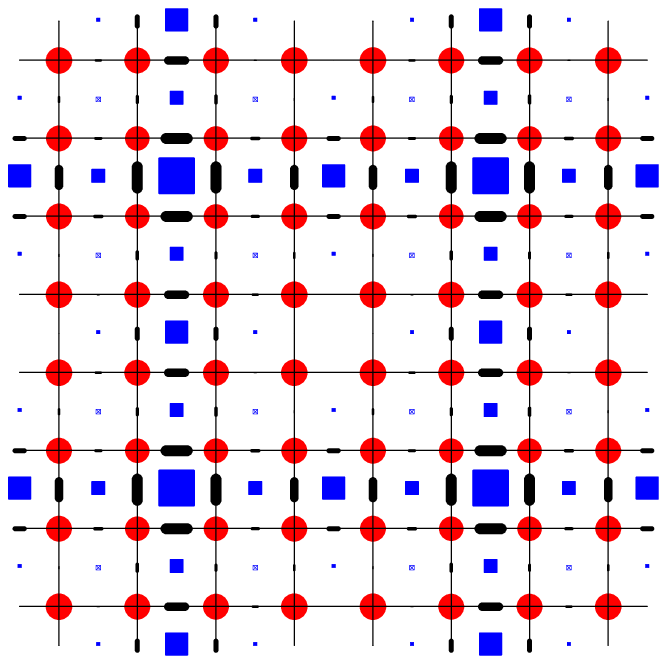}
\end{center}
\caption{Insulating state for doping $\delta=1/8$ with a $4 \times
4$ unit cell, with parameters as in
Figs.~\ref{fig:H=0}-\ref{fig:H=0.03125}. Now we are close to the
phase transition to the superfluid phase and the mean hole density
is more evenly distributed.} \label{fig:H=0.0625}
\end{figure}

In Fig~\ref{fig:phasediagram0}  (and
Figs~\ref{fig:H=0}-\ref{fig:H=0.0625}) we have indicated the links
of the direct lattice by solid lines. The size of the (black) bars
living on these links is proportional to the vortex kinetic energy
\begin{equation}
  \langle \psi_{a+\mu,\sigma}^{\ast} e^{2 \pi i (A_{a \mu}+\sigma C_{a\mu})}
\psi_{a\sigma} + \mbox{c.c.} \rangle \;,
\end{equation}
which we can physically interpret as the mean spin exchange energy
of the electrons on the direct lattice. The mean electron ring
exchange energy determined by
\begin{equation}
  \langle \psi_{a-}^{\ast} \psi_{a+} e^{2 \pi i \mathcal{Y}_a} + \mbox{c.c.} \rangle
\end{equation}
is depicted by (blue) boxes whose size is chosen to be
proportional to this quantity. If the exchange energy is negative
we have used a crossed box. Finally the (mean) density of holes on
the sites of the direct lattice is proportional to
\begin{equation}
  B_i \equiv \Delta_x A_{a_i y} - \Delta_y A_{a_i x} \;.
\end{equation}
For $H=0$ (zero doping) we expect this quantity to vanish, which
indeed is the case for the bond ordered or the plaquette state.
However, the CDW state in Fig~\ref{fig:phasediagram0} shows a
checkerboard pattern for $B_i$, which takes opposite values on the
two sublattices, and are depicted by (red) bullets or crossed
circles depending on whether $B_i$ is positive or negative.
Clearly this CDW state is not in a physically interesting regime
of couplings for the original dimer model as it applies to the
cuprates (the electron density is uniform in the Mott insulator at
half-filling); the lesson to be learned from this is that in our
model we cannot choose $g$ too large. For concreteness, let us
choose $g=2/e^2=50$ such that for sufficiently small $r$ we are
always in the columnar VBS state.

\begin{figure}[t]
\centering
\includegraphics[width=3in]{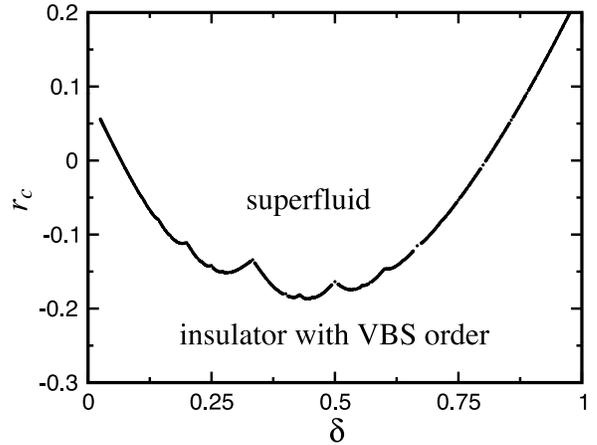}
\caption{Lower bound for the superfluid to insulator transition,
$r_c$ as a function of the hole density, $\delta$, in the
superfluid state ($\delta=2H$). There is a second order transition
at the points shown into an insulator with the same density. For
irrational $\delta$, this insulator is a floating Wigner solid.
The rational $\delta$ Mott insulators are expected to survive in
fixed density "Mott lobes" extending over a finite range of the
``chemical potential'' $H$, and these can pre-empt the transition
into the incommensurate floating solids. The re-entrance of the
curve above at large $H$ is in a regime of such large $\delta$
that we do not expect our small doping theory to remain valid.}
\label{fig:rcofH}
\end{figure}

To discuss doping holes into an insulator, we will start from a
columnar bond ordered state and follow its fate as we increase the
chemical potential $H$. First of all, it is instructive to
determine the critical value of $r$ at which we have the phase
transition from a superfluid to an insulating state as a function
of $H$. This transition occurs when the $\psi_{a \mu}$ condense,
and so a lower bound on the critical value of $r=r_c$ is obtained
by locating the point at which the lowest eigenvalue of the
quadratic part of the $\psi_{a \mu}$ free energy in
Eq.~(\ref{eq:F_MF}) turns negative. A graph of the lower bound of
$r_c (H)$ is shown in Fig~\ref{fig:rcofH}.
Determination of this lowest eigenvalue requires the Hofstadter
spectrum of $\psi_{a \mu}$, and so the graph in
Fig~\ref{fig:rcofH}) has a shape similar to the ground state
energy of the Hofstadter butterfly. For incommensurate values of
$H$, the transition at the points in Fig~\ref{fig:rcofH} is into a
floating Wigner solid with hole density $\delta = 2H$. We expect
that this will often be pre-empted by a first order transition
into a nearby commensurate Mott insulator with hole density
$\delta \neq 2H$, but instead at the rational value determined by
the Mott insulator. In the dual language of the `superconductor',
this Mott insulator has expelled some of the `magnetic flux' $H$.

In the following we will choose $r=-0.1$ such that we are in the
insulating state at zero doping, and enter the superfluid phase at
around $\delta \approx 0.15$. Our mean field theory also predicts
that at high enough doping concentration we are back in an
insulating state, but we do not expect our duality analysis to be
valid at such large $\delta$.
We do not make any pretense at completeness in the following phase
diagrams: they are merely representative structures obtained for a
sample set of parameters, and other parameters can clearly give a
wide variety of other phases. However, all such phases will obey
the general constraints we have delineated in I.

In Figs.~\ref{fig:H=0}-\ref{fig:H=0.0625} we show the states for
doping $\delta = 0,1/32,1/16$, and $1/8$. While at zero doping we
are in a columnar VBS state with vanishing hole density on the
sites, at doping $\delta=1/32$ we find a modified plaquette state
and which shows a concentration of the mean hole density at the
corners of one plaquette of its $8\times8$ unit cell. At the
center of this plaquette frustration leads to $\langle \psi_{a +}
\rangle = \langle \psi_{a -} \rangle = 0$. The variation of charge
density in the unit cell clearly results in a $2$-dimensional
charge density wave with wave-vectors ${\bf k}_x = (2\pi/8a,0)$
and ${\bf k}_y=(0,2\pi/8a)$ where $a$ is the lattice spacing. It
should also be noted that as for the plaquette state ($C$) in
Fig~\ref{fig:phasediagram0}, the state depicted in
Fig~\ref{fig:H=0.015625} is invariant under rotations by $90$
degrees about the plaquette with $\langle \psi_{a \sigma} \rangle
= 0$.

Doubling the doping concentration $\delta$, we find the state
depicted in Fig~\ref{fig:H=0.03125}, which is similar to the
previous state but has a unit cell of half its size and leads to a
modulation of the charge density with wave-vectors ${\bf k}_{\pm}
= (2\pi/8a,\pm 2\pi/8a)$. Finally at doping $\delta=1/8$, we find
the state depicted in Fig~\ref{fig:H=0.0625} which is also
invariant under rotations by $90$ degrees about a plaquette with
$\langle \psi_{a \sigma} \rangle = 0$ and has a $4\times4$ unit
cell. Since we are now close to the second order phase transition
to the superfluid phase, the mean hole density is now distributed
essentially uniformly over the lattice. However, there is still a
weak $2$-dimensional hole density modulation with wave-vectors
${\bf k}_x = (2\pi/4a,0)$ and ${\bf k}_y=(0,2\pi/4a)$

\subsection{Pairing symmetry}
\label{sec:dwave}

The analysis of this section has been carried out almost entirely
in a dual representation of the underlying electronic degrees of
freedom. While this has the benefit of properly accounting for the
non-perturbative dynamics of the compact U(1) gauge theory
$\mathcal{H}_{qd}$ in Eq.~(\ref{hqd}), it does obscure the
physical interpretation of the phases of the theory. We have
already shown in I how the density wave order can be thoroughly
characterized by the dual vortex degrees of freedom. In this
subsection we want to determine and analyze the corresponding dual
representation of correlators which characterize the
superconducting order.

We already discussed the issue of the pairing symmetry in
Section~\ref{sec:proj-real-symm} and found that certain key
determining factors could not be specified within the physics of
the dimer model alone. In Eq.~(\ref{eq:pfrequire}) we related the
electron pairing field to the product of microscopically
determined co-efficients $\mu_\alpha$ and degrees of freedom of
the quantum dimer model.
\begin{figure}
\centering
\includegraphics[width=2.5in]{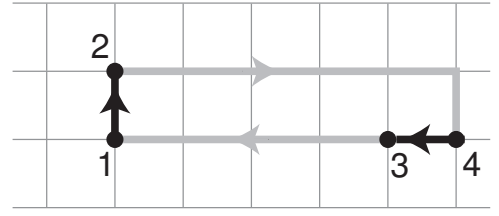}
\caption{Computation of the electron pair correlation function.
The sites 1,3 have $\eta_i = +1$, while 2,4 have $\eta_i=-1$. The
current $J_{31\mu}$ is equal to unity along the line connecting 3
to 1, and is zero elsewhere; other currents between two points are
defined similarly. We also define {\em loop} currents {\em e.g.}
$J_{1243\mu}$ is unity only along the loop around the points
1,2,4,3. An electron singlet pair is created on the link 1,2 and
annihilated on the link 3,4.} \label{figdwave}
\end{figure}
Let us consider the two-point correlation function for the pair field,
at two pairs of nearest-neighbor points $12$, $34$, where $2$ is
displaced in direction $\alpha$ from $1$, and $4$ is displaced in
direction $\beta$ from $3$ (see Fig~\ref{figdwave}\ for an example).
Then we can write the pairing correlator as
\begin{equation}
\langle \Psi_{12}^{\vphantom*} \Psi_{34}^\ast \rangle = \mu_\alpha \mu_\beta G_{12;34}
\label{eq:PsiG}
\end{equation}
where $G_{12;34}$ is a two-point correlator of the quantum dimer
model given by
\begin{equation}
G_{12;34} = \langle C^{\dagger}_{34} C_{12} \rangle\;, \label{p2}
\end{equation}
with the gauge-invariant hole pair annihilation operator
\begin{equation}
C_{12} = e^{i \eta_1\hat{\mathcal{A}}_{12}} e^{- i
\hat{\phi}_{1+}} e^{-i \hat{\phi}_{2-}}. \label{p1}
\end{equation}
We choose sites $1,3$ on the A sublattice ($\eta_i=+1$) and sites $2,4$
on the B sublattice ($\eta_i=-1$) for concreteness.

It now remains to evaluate $G_{12;34}$ within the context of duality
analysis of Section~\ref{sec:dimerdual}.  We will show that this
correlator factorizes in the limit of large separation between the
points 12 and 34, with no dependence on the relative orientations or
centers of 12 and 34.  This is indicative of superconducting
Off-Diagonal Long Range Order (ODLRO).  The symmetry of the pairing
state is then determined by the dependence of $\langle
\Psi_{12}^{\vphantom*}\Psi_{34}^* \rangle$ on either pair of neighboring
points $34$ or $12$, separately.  Consequently the pairing symmetry of
the superfluid ground state is determined by the $\mu_\alpha$ factors
controlling the particular PSG realized by the dimer model, {\sl and} by
the dependence of $G_{12;34}$ on $12$ or $34$ separately.  We will see
that, for the large $g$ limit on which we focus, $G_{12;34}$ is also
independence of the orientation or position of either pair of points
when they are well-separated.  Thus the pairing symmetry is determined
{\sl entirely} by the $\mu_\alpha$ factors.

The correlator $G_{12;34}$ can be computed in our dual
representation by inserting the operator in Eq.~(\ref{p1}), and
the corresponding operator at sites 3,4, into the transformations
leading to Eq.~(\ref{zdd}). After integrating over the degrees of
freedom on the direct lattice as before, the correlation function
is expressed as
\begin{eqnarray}
&& G_{12;34} = \frac{1}{\mathcal{Z}_{qd}} \sum_{\{J_{i\sigma\mu},
F_{a\mu}\}} \exp
\Biggl(- \frac{g}{2} \sum_{a}  F_{a \mu}^2 -  i \pi \sum_a F_{a\tau}  \nonumber \\
&&~~- \frac{1}{2 e^2} \sum_{i} \Bigr( \left[ J_{i + \mu} +
J_{31\mu}- (H+ \eta_i W ) \delta_{\mu\tau} \right]^2 \nonumber \\
&&+ \left[ J_{i-\mu} - J_{24\mu}- (H- \eta_i W ) \delta_{\mu\tau}
\right]^2\Bigr) \Biggr) \prod_{i\sigma} \delta \left( \Delta_\mu
J_{i\sigma\mu} \right) \nonumber
\\ &&~~~\times \prod_{a\mu} \delta \left( \epsilon_{\mu\nu\lambda}
\Delta_\nu F_{a \lambda} + J_{1243\mu}+\sum_\sigma \sigma
J_{i\sigma\mu} - \eta_i \delta_{\mu\tau} \right) \nonumber
\\
\label{p3}
\end{eqnarray}
Here the currents $J_{12\mu}$ etc. are fixed background currents
defined in Fig~\ref{figdwave} associated with the presence of the
source terms of the correlator. Notice that the delta function
constraints in Eq.~(\ref{p3}) differ importantly from those in
Eq.~(\ref{zdd}); consequently the solution of these constraints in
Eq.~(\ref{solvecons}) is replaced by
\begin{eqnarray}
J_{i \sigma \mu} &=& \epsilon_{\mu\nu\lambda} \Delta_\nu b_{a
\sigma \lambda} \nonumber \\
F_{a \mu} &=& \Delta_\mu N_a +\mathcal{B}_{a \mu} +
\mathcal{X}_{a\mu} - \sum_{\sigma} \sigma b_{a \sigma \mu},
\label{solvecons2}
\end{eqnarray}
where $\mathcal{B}_{a \mu}$ is a fixed background integer-valued
field which obeys
\begin{equation}
\epsilon_{\mu\nu\lambda} \Delta_\nu \mathcal{B}_{a \lambda} =
-J_{1243\mu}. \label{eq:curlB}
\end{equation}
Therefore, $\mathcal{B}_{a\mu}$ is the ``magnetic'' field
associated with the current loop in Fig~\ref{figdwave}a; a
convenient choice is to take only $\mathcal{B}_{a\tau}$ non-zero
along the links which pierce the current loop in
Fig~\ref{figdwave}a. Including the additional offset
$\mathcal{B}_{a\mu}$, we can continue the analysis of
Eq.~(\ref{p3}) as before, and obtain an expression corresponding
to that in Eq.~(\ref{zde}).

We now split the $b_{a\sigma\mu}$ gauge fields into its uniform
and staggered components as in Eq.~(\ref{bAC}). Manipulating the
expressions in the action in Eq.~(\ref{p3}) we find that they can
be split into those involving the uniform ($A_{a\mu}$) and
staggered ($C_{a\mu}$) components respectively. While the
resulting action has decoupled contributions from $A_{a \mu}$ and
$C_{a \mu}$, it should be kept in mind that their fluctuations are
not truly independent. In particular, because the $b_{a
\sigma\mu}$ must be integers, the $A_{a \mu}$ and $C_{a\mu}$ must
{\sl both} be either integers or half-integers. As we argue below,
the $A_{a\mu}$ field fluctuates strongly in the superfluid phase
(so that it can effectively be considered a continuous real
field), and so it appears reasonable to ignore such a constraint.
However it is possible we are overlooking some subtlety here, and
it may be worthwhile to re-examine this issue in future work.

For the uniform component we obtain the terms in the action
\begin{equation}
\mathcal{S}_A [J] = \frac{1}{4 e^2} \sum_a \left(
2\epsilon_{\mu\nu\lambda} \Delta_\nu A_{a \lambda} + J_{31\mu} +
J_{42\mu} - 2 H \delta_{\mu\tau} \right)^2\;. \label{eq:uniform}
\end{equation}
Notice that the source term here is the sum of the two currents
flowing from one pair field to the other, and so upon
coarse-graining it looks like the current of a charge 2 object
({\em i.e.} a Cooper pair) moving between 12 and 34. In the
superfluid phase, the uniform gauge field $A_{a\mu}$ is in its
Coulomb phase because the vortex fields $\psi_{a\sigma}$ are
gapped. So we can easily evaluate the expectation value of the
action in Eq.~(\ref{eq:uniform}) by treating $A_{a\mu}$ as a real
variable controlled by the usual Maxwell action. In this manner we
deduce that the contribution of the uniform component of the gauge
field to $G_{1234}$ is
\begin{eqnarray}
&& \frac{ \int \mathcal{D} A_{a\mu} e^{-\mathcal{S}_A [J]}}{\int
\mathcal{D} A_{a\mu} e^{-\mathcal{S}_A [0]}} = \label{eq:uniform2} \\
&& \exp \left ( \left( \Delta_\mu (J_{31\mu} + J_{42\mu}) \right)
\frac{1}{-4 e^2 \Delta_\lambda^2} \left( \Delta_\nu (J_{31\nu} +
J_{42\nu}) \right) \right) \;. \nonumber
\end{eqnarray}
The divergence of the currents is only non-zero at the location of
the Cooper pair source terms, and so we see that the correlator is
the magnetic energy of two `monopole' sources at these locations.
The square root of this corrrelator in the limit of infinite
separation is the superfluid order parameter, and this is given by
the finite self-energy of a single monopole. In addition, there is
a $1/r$ interaction between the monopoles, and this power-law
decay arises from the superflow fluctuations in the dual
superfluid.

We turn next to the contribution of the staggered field $C_{a\mu}$
to the action in Eq.~(\ref{p3}). In the gauge $N_a=0$, the
relevant contribution to the action is
\begin{eqnarray}
&& \mathcal{S}_C [J] = \frac{g}{2} \sum_a \left(\mathcal{B}_{a
\mu} + \mathcal{X}_{a\mu} - 2 C_{a\mu} \right)^2 \nonumber \\
&&~~~~~~~~+ i \pi \sum_a \left(\mathcal{B}_{a \tau} +
\mathcal{X}_{a\tau} - 2 C_{a\tau} \right) \label{eq:stagg} \\ && +
\frac{1}{4 e^2} \sum_a \left( 2\epsilon_{\mu\nu\lambda} \Delta_\nu
C_{a \lambda} + J_{31\mu} + J_{24\mu} - 2 \eta_i W
\delta_{\mu\tau} \right)^2\;. \nonumber
\end{eqnarray}
We now shift $2 C_{a\mu}$ by $\mathcal{B}_{a\mu} + \mathcal{X}_{a
\mu}$, and using Eqs.~(\ref{eq:curlX}) and (\ref{eq:curlB}) the
action above can be written as
\begin{eqnarray}
\mathcal{S}_C [J] &=& 2g \sum_a C_{a\mu}^2 - 2 i \pi \sum_a  C_{a\tau}  \nonumber \\
&+& \frac{1}{4 e^2} \sum_a \Bigl( 2\epsilon_{\mu\nu\lambda}
\Delta_\nu C_{a \lambda} - J_{12\mu} - J_{43\mu} \nonumber \\
&~&~~~~~~~~~- \eta_i (2W+1) \delta_{\mu\tau} \Bigr)^2\;.
\label{eq:stagg2}
\end{eqnarray}
The most important property of this expression is that it only
involves the local currents $J_{12\mu}$ and $J_{43\mu}$ which are
entirely localized at the Cooper pair source terms. So when these
sources move far apart, the correlator factorizes into two
contributions, one for each order parameter insertion.
Furthermore, because the staggered gauge correlations of
Eq.~(\ref{zde2}) decay exponentially with separation, it follows
that these insertions can be evaluated independently of each
other. As discussed earlier, this is indicative of ODLRO.

Next consider the dependence of $G_{12;34}$ upon $34$, keeping $12$
fixed and far away.  For two sets of points $34$ and $3'4'$, one has
\begin{equation}
  \label{eq:Grat}
  \frac{G_{12;34}}{G_{12;3'4'}} = \frac{\sum_{C_{a\mu}}
    e^{-S_C[J_{12},J_{34}]}}{\sum_{C_{a\mu}}
    e^{-S_C[J_{12},J_{3'4'}]}}
\end{equation}
For large $g$, one clearly sees that the numerator and denominator are
dominated by $C_{a\mu}=0$, independently of $J_{34}$ and $J_{3'4'}$.
Hence there is no directional or other dependence of $G_{12;34}$ on
either pair of points, and we can conclude that $G_{12;34}$ reaches a
featureless constant in the superfluid phase as 12 and 34 move
apart from each other.  The pairing signature in
Eq.~(\ref{eq:PsiG}) is determined by the $\mu_\alpha$, as promised.

\section{Collective $S=1$ `triplon' excitations}
\label{sec:triplon}

We discussed the inclusion of non-zero spin excitations in the
dimer model in Section~\ref{sec:spinful}, but focused entirely on
fermionic $S=1/2$ degrees of freedom. We also noted in
Section~\ref{sec:fermions} that neutral $S=1/2$ excitations did
not survive in the insulator: instead they confined in pairs to
create stable, bosonic $S=1$ `triplon' excitations. It is to this
triplon excitation that we now turn our attention to. In
principle, the triplon can also survive as a stable excitation
into the supersolid or superfluid phases, provided energy and
momentum conservation constraints prevent its decay into pairs of
$S=1/2$ Bogoliubov quasiparticles.

Microscopically, we can also see the necessity of including the
triplon excitation by referring to previous analyses using bond
operators \cite{ps,MVTU,GSU}. As discussed therein, two electrons
on sites at the ends of a square lattice link, in addition to
forming a spin singlet valence bond (a dimer), can also be in a
higher energy $S=1$ combination. It is the motion of this $S=1$
state that we wish to examine here.

Clearly, establishing the stability of the triplon requires an
accounting of fluctuations of the compact U(1) gauge field
$\hat{\mathcal{A}}$. So we need to work with dual fields of
Section~\ref{sec:dimerdual} to account for the $S=0$ sector. It is
cumbersome to carry triplon through this duality analysis, and so
we will attempt to guess its effective action using symmetry
considerations.

Our approach will be to combine the approach of
Ref.~\onlinecite{vs2} with the results of I and of
Section~\ref{sec:dimerdual}. As in Ref.~\onlinecite{vs2},  we
represent the triplon by a real vector field $\vec{n}_i$, where
the arrow denotes a vector in spin space. The spin operator on
site $j$ is related to $\vec{n}_j$ by
\begin{equation}
\vec{S}_j \propto \eta_i \vec{n}_j \;;\label{triplon1}
\end{equation}
so, on its own, $\vec{n}_i$ is a measure of spin correlations at
the commensurate antiferromagnetic wavevector $(\pi,\pi)$. Given
the short range antiferromagnetic couplings between the spins (or,
alternatively, using the results the microscopic calculation of
Ref.~\onlinecite{ps}), we follow Ref.~\onlinecite{vs2} in
postulating the following phenomenological quantum lattice model
for the $\vec{n}_i$:
\begin{eqnarray}
\mathcal{S}_{n0} &=& \int d \tau \sum_j \left[ \frac{1}{2} \left(
\frac{\partial \vec{n}_j }{\partial \tau} \right)^2 + \frac{s}{2}
\vec{n}_j^2 + \frac{u}{4} \left(
\vec{n}_j^2\right)^2 \right] \nonumber \\
&+& \int d \tau \sum_{\langle j j' \rangle} \frac{c^2}{2}
\left(\vec{n}_j - \vec{n}_{j'} \right)^2 \;. \label{triplon2}
\end{eqnarray}
Here $s$ is a parameter which determines the gap towards triplon
excitations, $u$ is a quartic non-linearity, and $c$ is a
velocity. In the absence of any coupling to the $S=0$ charged
excitations, Eqs.~(\ref{triplon1}), (\ref{triplon2}) predict that
the the lowest energy triplon excitation is at wavevector
$(\pi,\pi)$.

It is now useful to characterize universal aspects of the vortex
theory of Section~\ref{sec:dimerdual} using the methods of I. From
I we learn that for $f = \tilde{p}/\tilde{q}$ ($\tilde{p}$,
$\tilde{q}$ relatively prime integers), the low energy physics of
the $S=0$ sector is captured by $\tilde{q}$ vortex fields
$\varphi_\ell$, $\ell = 0,1 \ldots (\tilde{q}-1)$. The effective
action for the $\varphi_\ell$ fields is as described in I. From
these vortex fields we can also construct the density operators
\begin{equation}
\rho_{mn} =  \omega^{mn/2} \sum_{\ell = 0}^{\tilde{q}-1}
\varphi^{\ast}_\ell \varphi_{\ell+n} \omega^{\ell m},
\label{triplon3}
\end{equation}
where $m$, $n$ are integers, representing Fourier components of
`density'-like observables (such as pairing amplitude, exchange
energy, or site charge density) at wavevector $2 \pi f(m,n)$.
Finally, from the $\rho_{mn}$ we can Fourier transform to real
space and obtain $\delta \rho ({\bf r})$, the value the `density'
at the sites, links, and plaquettes of the square lattice; the
manner in which this is done, and the conventions for ${\bf r}$,
are described below and in Eq.~(2.26) of I.

We can now compute the effect of the $S=0$ sector on the triplon
$\vec{n}$ excitation just as in Ref.~\onlinecite{vs2}. We assume
that the variations in $\delta \rho ({\bf r})$ modulate the
exchange constants between (and the amplitudes of) the
$\vec{n}_j$, and so write down the simplest local couplings
between these degrees of freedom:
\begin{eqnarray}
\mathcal{S}_{n1} &=& \int d \tau \sum_j \Bigl[ \lambda_1
\delta\rho ({\bf r}_j ) \vec{n}_{j}^2 \nonumber \\
&+& \lambda_2 \sum_\alpha \delta \rho ({\bf r}_{j+\hat{e}_\alpha
/2})
\vec{n}_{j} \cdot \vec{n}_{j+\hat{e}_\alpha } \nonumber \\
&+& \lambda_3 \sum_\alpha \delta\rho ({\bf r}_j)
\vec{n}_{j-\hat{e}_\alpha} \cdot \vec{n}_{j+\hat{e}_\alpha} +
\cdots \Bigr] \label{triplon4}
\end{eqnarray}

We can now analyze $\mathcal{S}_{n0}+\mathcal{S}_{n1}$ as in
Refs.~\onlinecite{MVTU,GSU,vs2}. For static condensed
$\varphi_\ell$, as obtains in a commensurate Mott insulator, we
have a corresponding static modulation in $\delta \rho ({\bf r})$.
Its influence on the triplon excitation spectrum can be as
computed in Ref.~\onlinecite{vs2}. As was shown there, simple and
natural choices for the modulations agree with neutron scattering
observations \cite{jt04,hinkov,buyers,hayden} over a wide energy
range. We leave for future work the extension of these results to
the case where the $\varphi_\ell$ are fluctuating.

\section{Conclusions}
\label{sec:conc}

The primary purpose of this paper was to present a complete
treatment of the interplay between VBS order and superconductivity
in a model of a doped $S=1/2$ quantum antiferromagnet on the
square lattice. We aimed to do this in an approach which properly
accounted for strong gauge fluctuations even at non-zero hole
concentrations, $\delta$. While numerous previous studies have
studied the properties of effective gauge theories of quantum
antiferromagnets, essentially all have treated gauge fluctuations
only at $\delta=0$ (see however Ref.~\onlinecite{zlatko}). As we
recall briefly below, several key characteristics of our theory
were special to $\delta=0$.

Our analysis here was carried out in the context of the doped
quantum dimer model \cite{fradkiv}, which served as a convenient
effective theory for the spin $S=0$ sector. By a duality analysis
of this model in Section~\ref{sec:dimerdual}, we obtained a dual
theory of vortices in the local paired-electron superconducting
order. As in a previous paper \cite{psgbosons} (referred to as I)
on pure boson models, we found that these vortices appeared in
$\tilde{q}>1$ `flavors' determined by the electron-pair density in
a proximate Mott insulator with $\delta \neq 0$. There is much
evidence in the cuprates that they are proximate to a Mott
insulator with hole density $\delta=1/8$, and by
Eq.~(\ref{phieff}) this translates into a value $\tilde{q}=16$.
With the knowledge of $\tilde{q}$, we can then freely borrow over
essentially all of the results of I on the fluctuations of
superflow and charge/VBS order across superconducting, supersolid,
and insulating phases. The $\tilde{q}$ vortices transform under
projective representation of the square lattice space group (PSG)
which strongly constrains their effective action. Any
impurity-induced pinning of a vortex also induces VBS/charge order
by breaking the PSG among the $\tilde{q}$ vortices. We presented a
number of results on features of impurity induced pinning of
vortices in I, and the present paper has provided the promised
justification for their application to scanning tunnelling
microscopy (STM) experiments on the cuprates.
\cite{hoffman,fang,ali,mcelroy,hanaguri}

We also showed how the framework of the quantum dimer model could
be extended to include fermionic and non-zero spin excitations.
This led us to an unexpected, and fairly explicit, connection
between the dimer model and previous \cite{ba,sr,su2,su2long}
slave-particle U(1) and SU(2) gauge theories of quantum
antiferromagnets. This connection was established in the context
of mean-field saddle points in which all gauge fluctuations were
quenched. Our present analysis makes it quite clear that neglect
of such gauge fluctuations is dangerous at least at small
$\delta$: all the PSG properties of vortices described in the
previous paragraph emerged upon a careful treatment of such
fluctuations at $\delta \neq 0$.

The relationship between these different approaches to quantum
antiferromagnets is neatly highlighted by the key constraint
equation in Eq.~(\ref{constraint3})
\begin{equation}
\Delta_{\alpha} \hat{E}_{i\alpha} +
b^{\dagger}_{i+}b^{\vphantom\dagger}_{i+}-b^{\dagger}_{i-}b^{\vphantom\dagger}_{i-}
+ \sum_s d^{\dagger}_{is} d^{\vphantom\dagger}_{is}= 1,
\label{constraint3a}
\end{equation}
obeyed at every site of the lattice. Here $\hat{E}_{i \alpha}$ is
the integer-valued electric field of a compact U(1) gauge field
which is a measure of the number of singlet valence bonds on a
link of the direct square lattice. The $b_{i \pm}$ are bosons
representing holes which carry opposite gauge charges, and the
$d^{\dagger}_{is}$ are $S=1/2$ fermionic spinon degrees of
freedom. (It must be emphasized that this is merely a kinematic
description of theory, and the ultimate quantum numbers of the
emergent quasiparticles may well be different.) In the SU(2)
gauge-theoretic approaches of Refs~\onlinecite{su2,su2long}, the
first valence bond term in Eq.~(\ref{constraint3a}) is not
explicitly included, although it will be generated upon
renormalizing the theory to a lower energy scale; the constraint
is satisfied primarily by a half-filled band of $d$ fermions. In
our present dimer model approach, in contrast, the fluctuating
electric field term plays a central role in
Eq.~(\ref{constraint3a}); furthermore, we perform our duality
analysis about a fermionic vacuum in which there are two and zero
$d$ fermions (for an average of one) on the two sublattices
respectively (the PSG ensures that this choice of vacuum actually
breaks no symmetries). Alternatively stated in more physical
terms, the previous analyses \cite{su2,su2long} were carried out
about a background of fermions occupying a half-filled band of
single-particle states, while our duality analysis assumes that
the electrons are primarily in paired singlet valence bond states.
The consistency of our approach with the modulations observed in
STM studies of the vortex lattice \cite{hoffman} (discussed
above), and the absence (so far) of the staggered current patterns
near vortices predicted by the SU(2) gauge theory,\cite{su2vortex}
may tentatively be regarded as experimental evidence in support of
the approach advocated in I and the present paper.

While our duality analysis is best controlled when the fermionic $S=1/2$
quasiparticle excitations of the superconductor are fully gapped, the
structure of our theory does allow these quasiparticles to acquire a
gapless nodal structure in a superconductor, without a strong
qualitative impact on any of the phases of our theory.  Upon decreasing
$\delta$ from deep in the superconducting state, such nodal
quasiparticles can annihilate in pairs across a reduced Brillouin zone
boundary in a superconductor at $\delta=\delta_f$, as has been described
in previous work \cite{vs,ps,granath}.  Alternatively, one can imagine
that the nodal quasiparticles persist in the superconducting phase right
down to the Mott insulator-to-superconductor transition: the theory of I
will clearly not apply to such a transition, and obtaining the correct
theory for this case remains an important open problem.


\begin{acknowledgments}

We thank M.~P.~A.~Fisher, E.~Fradkin, S.~Kivelson, T.~Senthil, and
X.-G.~Wen for valuable discussions. In particular, conversations
with T.~Senthil stimulated us to discuss the relative
contributions of the electric field and fermion density to the
constraints in Eqs.~(\ref{constraint2}), (\ref{constraint3}),
(\ref{constraint4}), and (\ref{constraint3a}). This research was
supported by the National Science Foundation under grants
DMR-9985255 (L. Balents), DMR-0098226 (S.S.), and DMR-0210790,
PHY-9907949 at the Kavli Institute for Theoretical Physics (S.S.),
the Packard Foundation (L. Balents), the Deutsche
Forschungsgemeinschaft under grant BA 2263/1-1 (L. Bartosch), and
the John Simon Guggenheim Memorial Foundation (S.S.). S.S. thanks
the Aspen Center of Physics for hospitality. K.S. thanks
S.~M.~Girvin for support through ARO grant
1015164.2.J00113.627012.

\end{acknowledgments}

\appendix


\begin{thebibliography}{99}
\bibitem{fdfermion} The duality analysis of
Section~\ref{sec:dimerdual} is carried out with respect to the $f$
fermion vacuum, in that the $f$ fermion contribution to the l.h.s.
of Eq.~(\ref{constraint2}) is neglected. Note, however, that this
vaccuum does contain a large density of $d$ fermions: from
Eq.~(\ref{eq:spinons}), we see that there are two $d$ fermions on
every site of the $\eta_i=-1$ sublattice (for an average density
of one $d$ fermion per site), and this is an important
contribution to Eq.~(\ref{constraint3}). Although the $d$ fermion
contribution appears to break the translational symmetry of the
lattice, the PSG ensures that all physical observables of this
saddle point preserve all square lattice symmetries. The SU(2)
gauge theory saddle points of Refs.~\onlinecite{su2,su2long} also
have an average density of one $d$ fermion per site on the l.h.s.
of Eq.~(\ref{constraint4}), but this is realized in an explicitly
translationally invariant manner by a half-filled band of $d$
fermions.
\end{thebibliography}
\end{document}